\documentclass[aps,prd,nofootinbib,preprint,superscriptaddress]{revtex4-2}

\usepackage{amsmath, amssymb, amsthm, graphicx, epsfig, fancyhdr,epsfig, slashed}

\usepackage{tikzsymbols}
\usepackage{natbib}
\usepackage{float}

\usepackage{tikz,xcolor,hyperref}

\usepackage{mathtools}

\definecolor{lime}{HTML}{A6CE39}
\DeclareRobustCommand{\orcidicon}{
	\begin{tikzpicture}
	\draw[lime, fill=lime] (0,0) 
	circle [radius=0.2] 
	node[white] {{\fontfamily{qag}\selectfont \tiny ID}};
	\draw[white, fill=white] (-0.0625,0.095) 
	circle [radius=0.007];
	\end{tikzpicture}
	\hspace{-2mm}
}

\foreach \x in {A, ..., Z}{\expandafter\xdef\csname orcid\x\endcsname{\noexpand\href{https://orcid.org/\csname orcidauthor\x\endcsname}
			{\noexpand\orcidicon}}
}

\newcommand{\be}{\begin{equation}}
\newcommand{\ee}{\end{equation}}
\newcommand{\bea}{\begin{eqnarray}}
\newcommand{\eea}{\end{eqnarray}}
\newcommand{\MeV}{\mathop{\rm MeV}\nolimits}
\newcommand{\TeV}{\mathop{\rm TeV}\nolimits}

\newcommand{\eps}{\varepsilon}
\newcommand{\pfrac}[2]{\left(\frac{#1}{#2}\right)}

\newcommand{\sn}{\mathop{\rm sn}\nolimits}
\newcommand{\cn}{\mathop{\rm cn}\nolimits}
\newcommand{\dn}{\mathop{\rm dn}\nolimits}
\newcommand{\bbbone}{\hbox{\rm 1\kern-3pt l}}
\newcommand{\slp}{p\kern-5pt/}
\newcommand{\real}{\mathop{\rm Re}\nolimits}

\newcommand{\dAlem}{\hbox{\,\vbox{\hrule height0.3pt\hbox{\vrule width0.3pt
  \hbox{\vbox to8pt{\hbox to8pt{\hfill}\vfill}}\vrule width1.0pt}\hrule
  height1.0pt}\,}}
\newcommand{\dalem}{\hbox{\,\vbox{\hrule height0.2pt\hbox{\vrule width0.2pt
  \hbox{\vbox to5pt{\hbox to5pt{\hfill}\vfill}}\vrule width.6pt%
  \kern-0.1pt}\hrule height0.6pt}\,}}

\begin{document}

\title{Renormalisable Non-local Quark--Gluon Interaction:\\
\it{Mass Gap, Chiral Symmetry Breaking \& Scale Invariance}}

\author{Arpan Chatterjee\orcidD{}}
\email{arpan.chatterjee@ut.ee}
\affiliation{F\"u\"usika Instituut, Tartu Ulikool,
  W.~Ostwaldi 1, EE-50411 Tartu, Estonia}

\author{Marco Frasca\orcidA{}}
\email{marcofrasca@mclink.it}
\affiliation{Rome, Italy}

\author{Anish Ghoshal\orcidB{}}
\email{anish.ghoshal@roma2.infn.it}
\affiliation{Institute of Theoretical Physics, Faculty of Physics,
  University of Warsaw, ul.\ Pasteura 5, 02-093 Warsaw, Poland}

\author{Stefan Groote\orcidC{}}
\email{stefan.groote@ut.ee}
\affiliation{F\"u\"usika Instituut, Tartu Ulikool,
  W.~Ostwaldi 1, EE-50411 Tartu, Estonia}

\begin{abstract}
\textit{We derive a Nambu--Jona-Lasinio (NJL) model from a non-local gauge
theory and show that it has confining properties at low energies. In
particular, we present an extended approach to non-local QCD and a complete
revision of the technique of Bender, Milton and Savage applied to non-local
theories, providing a set of Dyson--Schwinger equations in differential form.
In the local case, we obtain closed form solutions in the simplest case of the
scalar field and extend it to the Yang--Mills field. In general, for
non-local theories, we use a perturbative technique and a Fourier series and
show how higher-order harmonics are heavily damped due to the presence of the
non-local factor. The spectrum of the theory is analysed for the non-local
Yang--Mills sector and found to be in agreement with the local results on the
lattice in the limit of the non-locality mass parameter running to infinity.
In the non-local case, we confine ourselves to a non-locality mass that is
sufficiently large compared to the mass scale arising from the integration of
the Dyson--Schwinger equations. Such a choice results in good agreement, in
the proper limit, with the spectrum of the local theory. We derive the gap
equation for the fermions in the theory that gives some indication of quark
confinement in the  non-local NJL case as well. Confinement seems to be a
rather ubiquitous effect that removes some degrees of freedom in the original
action, favouring the appearance of new observable states, as seen, e.g.,\ for
quantum chromodynamics at lower energies.}
\end{abstract}

\maketitle

\section{Introduction}

A lesson one can take from string theory is that strings are non-local
objects~\cite{Witten:1985cc,Freund:1987kt,Freund:1987ck,Brekke:1987ptq,%
Frampton:1988kr,Kostelecky:1988ta,Kostelecky:1989nt,Tseytlin:1995uq,%
Siegel:2003vt,Biswas:2004qu,Calcagni:2013eua,Calcagni:2014vxa}. Indeed, this
lesson is the original motivation for weakly non-local field theory to devise
a novel pathway for possible UV-regularised theories inspired by string
field theory~\cite{Witten:1985cc,Taylor:2003gn}, as investigated in several
Refs.\ since the 1990s~\cite{Moffat:1990jj,Evens:1990wf,Kleppe:1991rv,%
Tomboulis:1997gg,Clayton:2001xz,Biswas:2011ar,Modesto:2011kw,Modesto:2012ga,%
Modesto:2012ys,Modesto:2013oma,Biswas:2013cha,Modesto:2014xta,Modesto:2014lga,%
Talaganis:2014ida,Modesto:2015lna,Modesto:2015foa,Tomboulis:2015gfa,%
Modesto:2017hzl,Hashi:2018kag,Buoninfante:2018mre,Modesto:2021ief,%
Modesto:2021okr,Boos:2021chb,Boos:2021jih,Boos:2021lsj,Boos:2022biz}. Later
on, this pathway became an alternative to address the divergence and the
hierarchy problems in the Standard Model (SM) via generalising the kinetic
energy operators that are of second order in the derivative to an infinite
series of higher-order derivatives, suppressed by the scale $M^2$ of
non-locality~\cite{Krasnikov:1987yj,Moffat:1990jj,Biswas:2014yia}. Such
modifications of the kinetic energy sector in
introducing higher-order derivatives are free of
ghosts~\cite{Buoninfante:2018mre}~\footnote{The unitarity issues are well
addressed due to a certain prescription~\cite{Pius:2016jsl,Briscese:2018oyx,%
Chin:2018puw,Briscese:2021mob,Koshelev:2021orf}.} and also
cure the Higgs vacuum instability problem of the SM
Higgs~\cite{Ghoshal:2017egr}, as analysed by one of the authors. It was shown
that the $\beta$ function reaches a conformal limit resolving the Landau pole
issue in Quantum Field Theory~\cite{Ghoshal:2020lfd}. Therefore, by capturing
the infinite derivatives by an exponential of an entire function we obtain a
softened ultraviolet (UV) behaviour in the most suitable manner, without the
cost of introducing any new degrees of freedom that contribute to the particle
mass spectrum, since there are no new poles in the propagators of
infinite-derivative extensions.\footnote{For astrophysical implications,
dimensional transmutation and dark matter and dark energy phenomenology in
these theories see Refs.~\cite{Buoninfante:2018gce,Ghoshal:2018gpq,%
Frasca:2022vvp}.} A bound on the scale of non-locality from observations from
LHC and dark matter physics is $M\geq O(10)\TeV$~\cite{Biswas:2014yia,%
Ghoshal:2018gpq}. Moreover, the non-local theory leads to interesting
implications for proton decay and Grand Unified
Theories~\cite{Krasnikov:2020kgh}, as well as for braneworld
models~\cite{Nortier:2021six}. In addition, strongly coupled non-perturbative
regimes, exact $\beta$ functions and conditions of confinement in
higher-derivative non-local theories are being actively
investigated~\cite{Frasca:2020ojd,Frasca:2021iip,Frasca:2020jbe,%
Frasca:2022duz,Frasca:2022lwp,Frasca:2022pjf}. The results obtained so far
show that the effect of the non-locality in the strong coupling limits is to
dilute any mass gap (that may be present in the theory) in the UV regime the
system generates.\footnote{In the context of gravity theories one can get rid
of classical singularities, such as black hole
singularities~\cite{Biswas:2011ar,Biswas:2013cha,Frolov:2015bia,%
Frolov:2015usa,Koshelev:2018hpt,Koshelev:2017bxd,Buoninfante:2018xiw,%
Cornell:2017irh,Buoninfante:2018rlq,Buoninfante:2018stt,Abel:2019zou,%
Buoninfante:2020ctr} and cosmological singularities~\cite{Biswas:2005qr,%
Biswas:2006bs,Biswas:2010zk,Biswas:2012bp,Koshelev:2012qn,Koshelev:2018rau,%
Kolar:2021qox}. Recently, false vacuum tunnelling studies were carried out in
Ref.~\cite{Ghoshal:2022mnj}. Relations to string theory are found in
Refs.~\cite{Boos:2018bxf,Boos:2018bhd,Boos:2020kgj} (for an overview, cf.\
Ref.~\cite{Boos:2020qgg}).}

In order to render charge renormalisation finite, infinite-derivative terms
should be introduced for the fermion fields as well~\cite{Ghoshal:2020lfd}. An
analogous mechanism is the well-known Pauli--Villars regularisation scheme
where a mass dependence $\Lambda$ is introduced as the cut-off. The
infinite-derivative approach is to promote the Pauli--Villars cut-off
$\Lambda$ to the non-local energy scale $M$ of the theory. In the
infinite-derivative case, the non-local energy scale $M$ might play the role
of the ultraviolet cut-off. Recently, the infinite-derivative model for QED
and Yang--Mills has been reconsidered in view of its generalisation to the
SM~\cite{Ghoshal:2017egr,Ghoshal:2020lfd}.\footnote{See also
Refs.~\cite{Biswas:2014yia,Ghoshal:2018gpq} for a few phenomenological
applications to LHC physics and dark matter physics.} Indeed, this model leads
to a theory that is naturally free of quadratic divergencies, thus providing
an alternative way to the solution of the hierarchy
problem~\cite{Biswas:2014yia}. Higher-dimensional operators, containing new
interactions, naturally appear in higher-derivative theories with non-abelian
gauge structure. These operators soften quantum corrections in the UV regime
and extinguish divergencies in the radiative corrections. Nevertheless, as can
be easily understood by power-counting arguments, the new higher-dimensional
operators do not break renormalisability~\cite{Ghoshal:2020lfd}. This is due
to the improved ultraviolet behaviour of the bosonic propagator $P(k)$ in the
deep Euclidean region, which scales as $P(k)\sim\exp(-k^2/M^2)/k^4$, instead
of the usual propagator scaling in the local case as $P(k)\sim 1/k^2$ for
$k^2\to\infty$~\cite{Ghoshal:2020lfd}.

Note that the presence of the non-local energy scale $M$, associated with the
infinite-derivative term, manifestly breaks (at the classical level) the
conformal symmetry of the unbroken gauge sector. Therefore, one may wonder
whether this term can also trigger (dynamically) chiral symmetry breaking at
low energy, or in other words, whether the fermion field could dynamically
obtain a mass $m$ satisfying the condition $m<M$. The aim of the present paper
is to investigate this issue by analysing a general class of renormalisable
models containing infinite-derivative terms. In this paper, we will show that
a non-vanishing mass term for the fermion field can indeed be generated,
depending on the kind of interaction at hand, as a solution of the mass gap
equation. The fermion mass can be predicted, and it turns out to be a function
of the energy scale $M$. The effect of non-locality is to move a
possible violation of micro-causality to the region beyond the non-locality
mass scale $M$, making them possibly unobservable. On the other hand,
diagrammatic techniques {\it a la Feynman\/} cannot properly work in this
context, which makes our approach through solutions to the Dyson--Schwinger
set of equations more appealing. 

\textit{The paper is organised as follows:} In Sec.~\ref{SecID} we introduce
the infinite-derivative $SU(N)$. In Sec.~\ref{SecDSsol} we derive the set of
Dyson--Schwinger equations for non-local QCD. We solve these equations in a
perturbative manner by noting that higher harmonics are heavily damped by the
non-local factor. In Sec.~\ref{SecGB} the spectrum of the theory is analysed
for the non-local Yang--Mills sector and found to be in agreement with the
local results on lattice in the limit of the non-local scale running to
infinity. In the non-local case, we contend ourselves to have the non-local
scale sufficiently large with respect to the mass scale arising from the
integration of the Dyson--Schwinger equations. In Sec.~\ref{SecGF} we derive
the gap equation for the fermion in the theory and show that an identical
argument as given in Refs.~\cite{Frasca:2022lwp,Frasca:2022pjf} can be applied
here, giving some indication of quark confinement also in the non-local case.
Sec.~\ref{SecC} contains our conclusions. In Appendix~\ref{SecA} we derive the
Dyson--Schwinger equations for a $\phi^4$ theory in differential form. In
Appendix~\ref{SecB} we calculate the scalar two-point function.

\section{Infinite-derivative SU(N)\label{SecID}}
The introduction of infinite-derivative terms in the lagrangian is based of an
aproach by Lee and Wick~\cite{Lee:1969fy,Lee:1970iw}, cf.\ also
Ref.~\cite{Gabrielli:2007cp}. The Lee--Wick approach, understood as the
first terms in a series expansion, leads to a nonlocal infinite derivative
approach to $SU(N)$ gauge theories. In infinite-derivative $SU(N)$ with
massless fermions we start with the Lagrangian~\cite{Ghoshal:2020lfd}:
\begin{equation}\label{LprimeD}
{\cal L}'=\bar\psi\gamma^\mu(i\partial_\mu-g_sA_\mu^aT_a)\psi
  -\frac14F^a_{\mu\nu}e^{-f(D^2)}F_a^{\mu\nu}-\frac1{2\xi}(\partial_\mu
  A^\mu_a)e^{-f(D^2)}(\partial_\nu A^\nu_a)+j_a^\mu A^a_\mu
\end{equation}
where we assume the non-locality being only in the gauge sector and
$F^a_{\mu\nu}=\partial_\mu A^a_\nu-\partial_\nu A^a_\mu
-g_sf_{abc}A^b_\mu A^c_\nu$. Both the Yang--Mills term and the gauge-fixing
term are delocalised by the infinite-derivative exponential. A source term
$j_a^\mu A^a_\mu$ is added in order to use the Lagrangian for the generating
functional. As an example we can use $f(D^2)=D^2/M^2$ where
$D^{ab}_\mu=\delta^{ab}\partial_\mu-ig_sA^c_\mu(T_c)^{ab}$ is the covariant
derivative in the adjoint representation. For a large non-local scale $M^2$ it
has been shown in Refs.~\cite{Frasca:2020ojd,Frasca:2021iip} that $e^{f(D^2)}$
can be approximated by $e^{f(\dalem)}$, where $\dAlem=\partial^2$. Therefore,
one can start with
\begin{equation}
{\cal L}=\bar\psi\gamma^\mu(i\partial_\mu-g_sA_\mu^aT_a)\psi
  -\frac14F_{\mu\nu}^ae^{-f(\dalem)}F^{\mu\nu}_a
  -\frac1{2\xi}(\partial_\mu A^\mu_a)e^{-f(\dalem)}(\partial_\nu A^\nu_a)
  +j^\mu_aA_\mu^a.
\end{equation}
Dealing with the non-locality, we use a redefinition of fields by employing
\begin{equation}\label{redefA}
A^a_\mu=e^{\frac12f(\dalem)}\underline A^a_\mu
\end{equation}
which leads to
\begin{equation}\label{Linf}
{\cal L}=\bar\psi\gamma^\mu(i\partial_\mu-g_s[e^{\frac12f(\dalem)}\underline
  A^a_\mu]T_a)\psi+{\cal L}_{\rm YM+},
\end{equation}
where
\begin{equation}\label{LYM+}
  {\cal L}_{\rm YM+}=-\frac14\underline F^a_{\mu\nu}\underline F_a^{\mu\nu}
  -\frac1{2\xi}[\partial_\mu\underline A^\mu_a][\partial_\nu\underline A^\nu_a]
  +j_a^\mu[e^{\frac12f(\dalem)}\underline A^a_\mu].
\end{equation}
The underline stands for the field redefinition. Here and in the following,
the square brackets restrict the operational range of the differential
operators included. Using the fact that the Lagrangian is determined up to a
total divergence, one can use integration by parts to obtain
\begin{equation}\label{FFder}
F_{\mu\nu}^ae^{-f(\dalem)}F^{\mu\nu}_a
  =F_{\mu\nu}^ae^{-\frac12f(\dalem)}e^{-\frac12f(\dalem)}F^{\mu\nu}_a
  =[e^{-\frac12f(\dalem)}F_{\mu\nu}^a][e^{-\frac12f(\dalem)}F^{\mu\nu}_a]
  =:\underline F_{\mu\nu}^a\underline F^{\mu\nu}_a
\end{equation}
up to a total divergence (named $\mbox{div}$ in the following and vanishing
in the action integral), where $f[\dAlem g]=[\dAlem f]g+\mbox{div}$ and (by
induction) $f[\dAlem^ng]=[\dAlem^nf]g+\mbox{div}$ is used to share the
non-locality equally between the two field strength tensors. Explicitly, one
has
\begin{equation}
\underline F_{\mu\nu}^a=\partial_\mu\underline A_\nu^a
  -\partial_\nu\underline A_\mu^a+g_sf_{abc}e^{-\frac12f(\dalem)}
  [e^{\frac12f(\dalem)}\underline A_\mu^b]
  [e^{\frac12f(\dalem)}\underline A_\nu^c].
\end{equation}
For the modified Yang--Mills and gauge fixing Lagrangian one obtains
\begin{eqnarray}
{\cal L}_{\rm YM+}&=&-\frac14[\partial_\mu\underline A_\nu^a
  -\partial_\nu\underline A_\mu^a][\partial^\mu\underline A^\nu_a
  -\partial^\nu\underline A^\mu_a]
  -\frac1{2\xi}[\partial_\mu\underline A^\mu_a][\partial_\nu\underline A^\nu_a]
  +j^\mu_a[e^{\frac12f(\dalem)}\underline A_\mu^a]+\strut\nonumber\\&&\strut
  -\frac12g_sf_{abc}[\partial_\mu\underline A_\nu^a
  -\partial_\nu\underline A_\mu^a]
  e^{-\frac12f(\dalem)}[e^{\frac12f(\dalem)}\underline A^\mu_b]
  [e^{\frac12f(\dalem)}\underline A^\nu_c]+\strut\nonumber\\&&\strut
  -\frac14g_s^2f_{abc}f_{ade}[e^{-\frac12f(\dalem)}[e^{\frac12f(\dalem)}
  \underline A^\mu_b][e^{\frac12f(\dalem)}\underline A^\nu_c]]
  [e^{-\frac12f(\dalem)}[e^{\frac12f(\dalem)}\underline A_\mu^d]
  [e^{\frac12f(\dalem)}\underline A_\nu^e]]\ =\nonumber\\
  &=&-\frac12[\partial_\mu\underline A_\nu^a][\partial^\mu\underline A^\nu_a
  -\partial^\nu\underline A^\mu_a]-\frac1{2\xi}[\partial_\mu\underline A^\mu_a]
  [\partial_\nu\underline A^\nu_a]-ij^\mu_a[e^{\frac12f(\dalem)}\underline
  A_\mu^a]+\strut\nonumber\\&&\strut-\frac12g_sf_{abc}[\partial_\mu\underline
  A_\nu^a-\partial_\nu\underline A_\mu^a]
  e^{-\frac12f(\dalem)}[e^{\frac12f(\dalem)}\underline A^\mu_b]
  [e^{\frac12f(\dalem)}\underline A^\nu_c]+\strut\nonumber\\&&\strut
  -\frac14g_s^2f_{abc}f_{ade}[e^{\frac12f(\dalem)}\underline A^\mu_b]
  [e^{\frac12f(\dalem)}\underline A^\nu_c]e^{-f(\dalem)}
  [e^{\frac12f(\dalem)}\underline A_\mu^d]
  [e^{\frac12f(\dalem)}\underline A_\nu^e]+\mbox{div},
\end{eqnarray}
where the first term has been simplified and integrations by parts have been
used to move the exponential derivatives to the central position. This can
also be reverted if necessary, i.e., if the derivatives are inappropriate for
the variation of the fields. An ordinary integration by parts in the first
line for the second part of
${\cal S}=\int{\cal L}d^4x={\cal S}_f+{\cal S}_{\rm YM+}$ gives
\begin{eqnarray}
{\cal S}_{\rm YM+}&=&\int d^4x\Bigg[\frac12\underline A_\mu^a(x)
  (\dAlem\eta^{\mu\nu}-\partial^\mu\partial^\nu)\underline A_\nu^a(x)
  +\frac1{2\xi}\underline A_\mu^a(x)\partial^\mu\partial^\nu\underline
  A_\nu^a(x)+j^\mu_a(x)[e^{\frac12f(\dalem)}\underline A_\mu^a(x)]
  +\strut\kern-1pt\nonumber\\&&\strut
  -\frac12g_sf_{abc}[\partial_\mu\underline A_\nu^a(x)
  -\partial_\nu\underline A_\mu^a(x)]
  e^{-\frac12f(\dalem)}[e^{\frac12f(\dalem)}\underline A^\mu_b(x)]
  [e^{\frac12f(\dalem)}\underline A^\nu_c(x)]+\strut\nonumber\\&&\strut
  -\frac14g_s^2f_{abc}f_{cde}[e^{\frac12f(\dalem)}\underline A^\mu_a(x)]
  [e^{\frac12f(\dalem)}\underline A^\nu_b(x)]e^{-f(\dalem)}
  [e^{\frac12f(\dalem)}\underline A_\mu^d(x)]
  [e^{\frac12f(\dalem)}\underline A_\nu^e(x)]\Bigg].
\end{eqnarray}
The Euler--Lagrange equation $\delta{\cal S}/\delta\underline A_\mu^a=0$ for
the Yang--Mills field reads
\begin{eqnarray}\label{EL}
\lefteqn{g_se^{\frac12f(\dalem)}\bar\psi\gamma^\mu T_a\psi
  \ =\ (\dAlem\eta^{\mu\nu}-\partial^\mu\partial^\nu)\underline A_\nu^a
  +\frac1\xi\partial^\mu\partial^\nu\underline A_\nu^a
  +e^{\frac12f(\dalem)}j^\mu_a+\strut}\nonumber\\&&\strut
  -g_sf_{abc}e^{-\frac12f(\dalem)}\partial_\nu[e^{\frac12f(\dalem)}
  \underline A^\mu_b][e^{\frac12f(\dalem)}\underline A^\nu_c]
  -g_sf_{abc}e^{\frac12f(\dalem)}[e^{\frac12f(\dalem)}\underline A_\nu^b]
  e^{-\frac12f(\dalem)}(\partial^\mu\underline A^\nu_c
  -\partial^\nu\underline A^\mu_c)+\kern-12pt\strut\nonumber\\[7pt]&&\strut
  -g_s^2f_{abc}f_{cde}e^{\frac12f(\dalem)}[e^{\frac12f(\dalem)}
  \underline A_\nu^b]e^{-f(\dalem)}[e^{\frac12f(\dalem)}\underline A^\mu_d]
  [e^{\frac12f(\dalem)}\underline A^\nu_e].
\end{eqnarray}
In the following we use Feynman gauge $\xi=1$ to further simplify the first
line. Applying the formalism of Bender, Milton and
Savage~\cite{Bender:1999ek} to the Euler--Lagrange equation~(\ref{EL}) leads
to the tower of Dyson--Schwinger equations which is formulated in terms of
multiple-point Green functions. This approach is displayed in Appendices~A
and~B for a local $\phi^4$ theory.

\section{Solution of the Dyson--Schwinger equations\label{SecDSsol}}
The creation of the tower of Dyson--Schwinger equations is explained in detail
in Appendix~B of Ref.~\cite{Frasca:2020ojd}. We do not repeat it here to avoid
reprinting previously published material. Instead, we give some guidelines. To
start with, one takes the expectation value of the Euler--Lagrange
equation~(\ref{EL}), weighted by the generating functional
\begin{equation}
Z[j]=\int[d\underline{A}]\exp\left(i\int{\cal L}_{\rm YM+}d^4x\right),
\end{equation}
where the current $j$ is already contained in ${\cal L}_{\rm YM+}$, indicated
by the plus sign in the index. One has
$\langle e^{\frac12f(\dalem_x)}\underline A_\mu^a(x)\rangle
=:Z[j]e^{\frac12f(\dalem_x)}\underline G_{1\mu}^{(j)a}(x)$. Applying either
additional partial derivative like
$\langle e^{\frac12f(\dalem_x)}\partial^x_\nu\underline A_\mu^a(x)\rangle
  =Z[j]e^{\frac12f(\dalem_x)}\partial^x_\nu\underline G_{1\mu}^{(j)a}(x)$,
or calculating a variation w.r.t.\ e.g.\ $j^\nu_b(x')$,
\begin{equation}\label{G1G1}
\langle e^{\frac12f(\dalem_x)}\underline A_\mu^a(x)
  e^{\frac12f(\dalem_{x'})}\underline A_\nu^b(x')\rangle
  =Z[j]e^{\frac12f(\dalem_x)}\underline G_{2\mu\nu}^{(j)ab}(x,x')
  +Z[j]e^{\frac12f(\dalem_x)}\underline G_{1\mu}^{(j)a}(x)
  e^{\frac12f(\dalem_{x'})}\underline G_{1\nu}^{(j)b}(x'),\kern-12pt
\end{equation}
one finally obtains
\begin{eqnarray}\label{EL2}
\lefteqn{\dAlem\underline G_{1\mu}^{(j)a}(x)+e^{\frac12f(\dalem)}j_\mu^a(x)
  \ =}\nonumber\\
  &=&g_sf_{abc}\Big\{e^{-\frac12f(\dalem)}\partial^\nu
  \Big([e^{\frac12f(\dalem)}\underline G_{2\mu\nu}^{(j)bc}(x,x)]
  +[e^{\frac12f(\dalem)}\underline G_{1\mu}^{(j)b}(x)]
  [e^{\frac12f(\dalem)}\underline G_{1\nu}^{(j)c}(x)]\Big)
  +\strut\nonumber\\&&\strut\qquad\qquad
  +e^{-\frac12f(\dalem)}[e^{\frac12f(\dalem)}(\partial_\mu\underline
  G_{2\nu b}^{(j)c\nu}(x,x)-\partial_\nu\underline G_{2\mu b}^{(j)c\nu}(x,x))]
  +\strut\nonumber\\&&\strut\qquad\qquad
  +e^{\frac12f(\dalem)}[e^{\frac12f(\dalem)}\underline G_{1b}^{(j)\nu}(x)]
  e^{-f(\dalem)}[e^{\frac12f(\dalem)}(\partial_\mu\underline G_{1\nu}^{(j)c}
  (x)-\partial_\nu\underline G_{1\mu}^{(j)c}(x))]\Big\}
  +\strut\nonumber\\&&\strut+g_s^2f_{abc}f_{cde}\Big\{e^{-\frac12f(\dalem)}
  \Big([e^{\frac12f(\dalem)}\underline G_{3\mu\nu b}^{(j)de\nu}(x,x,x)]
  +[e^{\frac12f(\dalem)}\underline G_{2\mu b}^{(j)d\nu}(x,x)]
  [e^{\frac12f(\dalem)}\underline G_{1\nu}^{(j)e}(x)]
  +\strut\nonumber\\&&\strut\qquad\qquad\qquad\qquad
  +[e^{\frac12f(\dalem)}\underline G_{1\mu}^{(j)d}(x)]
  [e^{\frac12f(\dalem)}\underline G_{2\nu b}^{(j)e\nu}(x,x)]\Big)
  +\strut\nonumber\\&&\strut\qquad\qquad
  +e^{\frac12f(\dalem)}[e^{\frac12f(\dalem)}\underline G_{1b}^{(j)\nu}(x)]
  e^{-f(\dalem)}\Big([e^{\frac12f(\dalem)}\underline G_{2\mu\nu}^{(j)de}(x,x)]
  +\strut\nonumber\\&&\strut\qquad\qquad\qquad\qquad\qquad\qquad\qquad
  +[e^{\frac12f(\dalem)}\underline G_{1\mu}^{(j)d}(x)]
  [e^{\frac12f(\dalem)}\underline G_{1\nu}^{(j)e}(x)]\Big)\Big\}.
\end{eqnarray}
The equation of motion for the two-point function is obtained by variation
w.r.t.\ $j^\lambda_h(y)$,
\begin{eqnarray}
\lefteqn{\dAlem\underline G_{2\mu\lambda}^{(j)ah}(x,y)-i\delta^{ah}
  \eta_{\mu\lambda}e^{\frac12f(\dalem)}\delta^{(4)}(x-y)\ =}\nonumber\\
  &=&g_sf_{abc}\Big\{e^{-\frac12f(\dalem)}\partial^\nu
  \Big([e^{\frac12f(\dalem)}\underline G_{3\mu\nu\lambda}^{(j)bch}(x,x,y)]
  +[e^{\frac12f(\dalem)}\underline G_{2\mu\lambda}^{(j)bh}(x,y)]
  [e^{\frac12f(\dalem)}\underline G_{1\nu}^{(j)c}(x)]
  +\strut\nonumber\\&&\strut\qquad\qquad
  +[e^{\frac12f(\dalem)}\underline G_{1\mu}^{(j)b}(x)]
  [e^{\frac12f(\dalem)}\underline G_{2\nu\lambda}^{(j)ch}(x,y)]\Big)
  +\strut\nonumber\\&&\strut\qquad
  +e^{-\frac12f(\dalem)}[e^{\frac12f(\dalem)}(\partial_\mu\underline
  G_{3\nu b\lambda}^{(j)c\nu h}(x,x,y)-\partial_\nu\underline
  G_{3\mu b\lambda}^{(j)c\nu h}(x,x,y))]+\strut\nonumber\\&&\strut\qquad
  +e^{\frac12f(\dalem)}\Big(
  [e^{\frac12f(\dalem)}\underline G_{2b\lambda}^{(j)\nu h}(x,y)]e^{-f(\dalem)}
  [e^{\frac12f(\dalem)}(\partial_\mu\underline G_{1\nu}^{(j)c}(x)
  -\partial_\nu\underline G_{1\mu}^{(j)c}(x))]
  +\strut\nonumber\\&&\strut\qquad\qquad
  +[e^{\frac12f(\dalem)}\underline G_{1b}^{(j)\nu}(x)]
  e^{-f(\dalem)}[e^{\frac12f(\dalem)}(\partial_\mu\underline
  G_{2\nu\lambda}^{(j)ch}(x,y)-\partial_\nu\underline G_{2\mu\lambda}^{(j)ch}
  (x,y))]\Big)\Big\}+\strut\nonumber\\&&\strut
  +g_s^2f_{abc}f_{cde}\Big\{e^{-\frac12f(\dalem)}\Big([e^{\frac12f(\dalem)}
  \underline G_{4\mu\nu b\lambda}^{(j)de\nu h}(x,x,x,y)]
  +\strut\nonumber\\&&\strut\qquad\qquad
  +[e^{\frac12f(\dalem)}\underline G_{3\mu\nu b}^{(j)d\nu h}(x,x,y)]
  [e^{\frac12f(\dalem)}\underline G_{1\nu}^{(j)e}(x)]
  +[e^{\frac12f(\dalem)}\underline G_{2\mu b}^{(j)d\nu}(x,x)]
  [e^{\frac12f(\dalem)}\underline G_{2\nu\lambda}^{(j)eh}(x,y)]
  +\strut\kern-25pt\nonumber\\&&\strut\qquad\qquad
  +[e^{\frac12f(\dalem)}\underline G_{2\mu\lambda}^{(j)dh}(x,y)]
  [e^{\frac12f(\dalem)}\underline G_{2\nu b}^{(j)e\nu}(x,x)]
  +[e^{\frac12f(\dalem)}\underline G_{1\mu}^{(j)d}(x)]
  [e^{\frac12f(\dalem)}\underline G_{3\nu b\lambda}^{(j)e\nu h}(x,x,y)]\Big)
  +\strut\kern-32pt\nonumber\\&&\strut\qquad
  +e^{\frac12f(\dalem)}
  [e^{\frac12f(\dalem)}\underline G_{2b\lambda}^{(j)\nu h}(x,y)]
  \times\strut\nonumber\\&&\strut\qquad\qquad\times
  e^{-f(\dalem)}([e^{\frac12f(\dalem)}\underline G_{2\mu\nu}^{(j)de}(x,x)]
  +[e^{\frac12f(\dalem)}\underline G_{1\mu}^{(j)d}(x)]
  [e^{\frac12f(\dalem)}\underline G_{1\nu}^{(j)e}(x)])
  +\strut\nonumber\\&&\strut\qquad
  +e^{\frac12f(\dalem)}[e^{\frac12f(\dalem)}\underline G_{1b}^{(j)\nu}(x)]
  e^{-f(\dalem)}\Big([e^{\frac12f(\dalem)}\underline
  G_{3\mu\nu\lambda}^{(j)deh}(x,x,y)]+\strut\\&&\strut\qquad\qquad
  +[e^{\frac12f(\dalem)}\underline G_{2\mu\lambda}^{(j)dh}(x,y)]
  [e^{\frac12f(\dalem)}\underline G_{1\nu}^{(j)e}(x)]
  +[e^{\frac12f(\dalem)}\underline G_{1\mu}^{(j)d}(x)]
  [e^{\frac12f(\dalem)}\underline G_{2\nu\lambda}^{(j)eh}(x,y)]\Big)\Big\},
  \nonumber
\end{eqnarray}
with $\delta\underline G_{1\mu}^{(j)a}(x)/\delta j^\lambda_h(y)
  =\underline G_{2\mu\lambda}^{(j)ah}(x,y)$. The system of equations of motion
for the complete set of components of the Green functions for the Yang--Mills
field cannot be treated exactly. Instead, we use a mapping to the scalar case.
The mapping theorem introduced in Refs.~\cite{Frasca:2007uz,Frasca:2009yp} is
based on Andrei Smilga's solution for the problem that ``the Yang--Mills
system is not exactly solvable, \dots\ in contrast to \dots\ some early
hopes'' (cf.\ Sec.~1.2 in Ref.~\cite{Smilga:2001ck}). Indeed, for more than
one independent component, in solving the system of equations one observes
chaotic behaviour (cf.\ Refs.~\cite{Matinyan:1981ys,Savvidy:1982wx,%
  Savvidy:1982jk}). Even though the importance of such observations for the
macroscopic behaviour is questionable, the safe path is to use a mapping to a
single scalar function that we dub $\underline\phi(x)$, representing an exact
solution for the 1P-correlation function in the scalar case. This is realized
here by using $\underline G_{1\mu}^{(0)a}(x)=\eta_\mu^a\underline\phi(x)$ and
$\underline G_{2\mu\nu}^{(0)ab}(x,y)=\eta_{\mu\nu}\delta^{ab}\underline
G_2(x-y)$ for $j_\mu^a(x)$ set to zero, where $\eta_\mu^a$ are the components
of the polarisation vector and $\eta_{\mu\nu}$ are the components of the
Minkowski metric. The two-point Green functions from $x$ to $x$ become
constants with vanishing derivative while $n$-point Green functions with $n>2$
can be set to zero if at least two arguments coincide. One obtains
\begin{eqnarray}
\lefteqn{\eta_\mu^a\dAlem\underline\phi(x)
  \ =\ g_sf_{abc}\Big\{e^{-\frac12f(\dalem)}\partial^\nu\eta_\mu^b\eta_\nu^c
  [e^{\frac12f(\dalem)}\underline\phi(x)]^2+\strut}\nonumber\\&&\strut\qquad
  +e^{\frac12f(\dalem)}\eta^\nu_b[e^{\frac12f(\dalem)}\underline\phi(x)]
  e^{-f(\dalem)}[e^{\frac12f(\dalem)}(\eta_\nu^c\partial_\mu\underline\phi(x)
  -\eta_\mu^c\partial_\nu\underline\phi(x))]\Big\}+\strut\nonumber\\&&\strut
  +g_s^2f_{abc}f_{cde}\Big\{e^{-\frac12f(\dalem)}(\eta_\mu^\nu\delta_b^d
  \eta_\nu^e+\eta_\nu^\nu\delta_b^e\eta_\mu^d)[e^{\frac12f(\dalem)}
  \underline G_2(x-x)][e^{\frac12f(\dalem)}\underline\phi(x)]
  +\strut\nonumber\\&&\strut\qquad
  +e^{\frac12f(\dalem)}\eta^\nu_b[e^{\frac12f(\dalem)}\underline\phi(x)]
  e^{-f(\dalem)}\Big(\eta_{\mu\nu}\delta^{de}[e^{\frac12f(\dalem)}\underline
  G_2(x-x)]+\eta_\mu^d\eta_\nu^e[e^{\frac12f(\dalem)}\underline\phi(x)]^2\Big)
  \Big\}\ =\nonumber\\
  &=&g_sf_{abc}\Big\{e^{-\frac12f(\dalem)}\eta_\mu^b\eta^\nu_c\partial_\nu
  [e^{\frac12f(\dalem)}\underline\phi(x)]^2+\strut\nonumber\\&&\strut\qquad
  +e^{\frac12f(\dalem)}[e^{\frac12f(\dalem)}\underline\phi(x)]
  e^{-f(\dalem)}[e^{\frac12f(\dalem)}(\eta^\nu_b\eta_\nu^c\partial_\mu\underline
  \phi(x)-\eta^\nu_b\eta_\mu^c\partial_\nu\underline\phi(x))]\Big\}
  +\strut\nonumber\\&&\strut
  +g_s^2f_{abc}f_{cde}\Big\{(\delta^{bd}\eta_\mu^e+D\delta^{be}\eta_\mu^d)
  e^{-\frac12f(\dalem)}[e^{\frac12f(\dalem)}\underline G_2(x-x)]
  [e^{\frac12f(\dalem)}\underline\phi(x)]+\strut\nonumber\\&&\strut\qquad\qquad
  +\eta^\nu_b\eta_\mu^d\eta_\nu^ee^{\frac12f(\dalem)}[e^{\frac12f(\dalem)}
  \underline\phi(x)]e^{-f(\dalem)}[e^{\frac12f(\dalem)}\underline\phi(x)]^2
  \Big\}
\end{eqnarray}
and
\begin{eqnarray}
\lefteqn{\delta^{ah}\eta_{\mu\lambda}\dAlem\underline G_2(x-y)-i\delta^{ah}
  \eta_{\mu\lambda}e^{\frac12f(\dalem)}\delta^{(4)}(x-y)\ =}\nonumber\\
  &=&g_sf_{abc}\Big\{e^{-\frac12f(\dalem)}(\eta_{\mu\lambda}\delta^{bh}
  \eta_\nu^c+\eta_{\nu\lambda}\eta_\mu^b\delta^{ch})\partial^\nu
  [e^{\frac12f(\dalem)}\underline\phi(x)][e^{\frac12f(\dalem)}
  \underline G_2(x-y)]+\strut\nonumber\\&&\strut\qquad
  +e^{\frac12f(\dalem)}\delta_b^h\eta_\lambda^\nu[e^{\frac12f(\dalem)}
  \underline G_2(x-y)]e^{-f(\dalem)}[e^{\frac12f(\dalem)}(\eta_\nu^c
  \partial_\mu\underline\phi(x)-\eta_\mu^c\partial_\nu\underline\phi(x))]
  +\strut\nonumber\\&&\strut\qquad
  +e^{\frac12f(\dalem)}\eta^\nu_b\delta^{ch}[e^{\frac12f(\dalem)}
  \underline\phi(x)]e^{-f(\dalem)}[e^{\frac12f(\dalem)}
  (\eta_{\nu\lambda}\partial_\mu-\eta_{\mu\lambda}\partial_\nu)
  \underline G_2(x-y)]+\strut\nonumber\\&&\strut
  +g_s^2f_{abc}f_{cde}\Big\{\eta_{\mu\lambda}(\delta_b^d\delta^{eh}
  +\delta_b^e\delta^{dh}\eta_\nu^\nu)[e^{\frac12f(\dalem)}\underline G_2(x-x)]
  [e^{\frac12f(\dalem)}\underline G_2(x-y)]+\strut\nonumber\\&&\strut\qquad
  +\eta_\lambda^\nu\delta_b^he^{\frac12f(\dalem)}[e^{\frac12f(\dalem)}
  \underline G_2(x-y)]e^{-f(\dalem)}(\eta_{\mu\nu}\delta^{de}
  [e^{\frac12f(\dalem)}\underline G_2(x-x)]
  +\eta_\mu^d\eta_\nu^e[e^{\frac12f(\dalem)}\underline\phi(x)]^2)
  +\strut\kern-15pt\nonumber\\&&\strut\qquad
  +\eta^\nu_b(\eta_{\mu\lambda}\delta^{dh}\eta_\nu^e+\eta_{\nu\lambda}
  \delta^{eh}\eta_\mu^d)[e^{\frac12f(\dalem)}\underline G_2(x-y)]
  [e^{\frac12f(\dalem)}\underline\phi(x)]^2\Big\}.
\end{eqnarray}
Contracting with $\eta^\mu_a$ and $\delta_{ah}\eta^{\mu\lambda}$ and using the
orthogonality and completeness relations
\begin{equation}\label{complete}
\eta_\mu^a\eta^\mu_b=-\delta^a_b,\qquad
\eta_\mu^a\eta_\nu^a=-(N_c^2-1)\eta_{\mu\nu}/D,
\end{equation}
where $(\eta_{\mu\nu})=\mbox{diag}(1,-1,-1,-1)$
($D=\eta_{\mu\lambda}\eta^{\mu\lambda}$ is the space--time dimension), one
obtains
\begin{eqnarray}
-(N_c^2-1)\dAlem\underline\phi(x)
  &=&g_s^2f_{abc}f_{cde}\Big\{-(\delta_b^d\delta_a^e+D\delta_b^e\delta_a^d)
  e^{-\frac12f(\dalem)}[e^{\frac12f(\dalem)}\underline G_2(x-x)]
  [e^{\frac12f(\dalem)}\underline\phi(x)]+\strut\nonumber\\&&\strut\qquad\qquad
  +\delta_b^e\delta_a^de^{\frac12f(\dalem)}[e^{\frac12f(\dalem)}
  \underline\phi(x)]e^{-f(\dalem)}[e^{\frac12f(\dalem)}\underline\phi(x)]^2
  \Big\}\ =\nonumber\\
  &=&N_c(N_c^2-1)g_s^2\Big\{-(D-1)e^{-\frac12f(\dalem)}[e^{\frac12f(\dalem)}
  \underline G_2(x-x)][e^{\frac12f(\dalem)}\underline\phi(x)]
  +\strut\nonumber\\&&\strut\qquad\qquad
  +e^{\frac12f(\dalem)}[e^{\frac12f(\dalem)}\underline\phi(x)]e^{-f(\dalem)}
  [e^{\frac12f(\dalem)}\underline\phi(x)]^2\Big\}
\end{eqnarray}
and
\begin{eqnarray}
\lefteqn{D(N_c^2-1)\left(\dAlem\underline G_2(x-y)-ie^{\frac12f(\dalem)}
  \delta^{(4)}(x-y)\right)\ =}\nonumber\\
  &=&g_s^2f_{abc}f_{cde}\Big\{D(\delta_b^d\delta_a^e+D\delta_b^e\delta_a^d)
  [e^{\frac12f(\dalem)}\underline G_2(x-x)][e^{\frac12f(\dalem)}\underline
  G_2(x-y)]+\strut\nonumber\\&&\strut\qquad
  -(D\delta_a^d\delta_b^e+\delta_a^e\delta_b^d)[e^{\frac12f(\dalem)}
  \underline G_2(x-y)][e^{\frac12f(\dalem)}\underline\phi(x)]^2\Big\}
  \ =\nonumber\\
  &=&(D-1)N_c(N_c^2-1)g_s^2\Big\{D[e^{\frac12f(\dalem)}\underline G_2(x-x)]
  -[e^{\frac12f(\dalem)}\underline\phi(x)]^2\Big\}
  [e^{\frac12f(\dalem)}\underline G_2(x-y)],
\end{eqnarray}
where $f_{abc}f_{abd}=N_c\delta_{cd}$ for two and, therefore,
$f_{abc}f_{abc}=N_c(N_c^2-1)$ for three summed indices have been used. The
common factor cancels, and one obtains
\begin{equation}\label{G1eq}
(\dAlem+\Delta m_G^2)\underline\phi(x)+\lambda
  e^{\frac12f(\dalem)}[e^{\frac12f(\dalem)}\underline\phi(x)]e^{-f(\dalem)}
  [e^{\frac12f(\dalem)}\underline\phi(x)]^2=0
\end{equation}
and
\begin{equation}\label{G2eq}
(\dAlem+\Delta m_G^2)\underline G_2(x-y)+\frac{D-1}D\lambda
  [e^{\frac12f(\dalem)}\underline\phi(x)]^2
  [e^{\frac12f(\dalem)}\underline G_2(x-y)]
  =ie^{\frac12f(\dalem)}\delta^{(4)}(x-y),
\end{equation}
where $\lambda=N_cg_s^2$ and
$\Delta m_G^2=-(D-1)e^{\frac12f(\dalem_x)}\lambda G_2(x-x')|_{x'=x}$.

Looking at Eq.~(\ref{G1eq}), the solution for the corresponding local
one-point function, given by $\underline\phi_0(x)=\hat\phi_0(kx)=\mu
\sn(kx+\theta|\kappa)$ \cite{Frasca:2007uz}, obeys the dispersion relation
$k^2=\Delta m_G^2+\lambda\mu^2/2$ with $\kappa=(\Delta m_G^2-k^2)/k^2$ and
$\theta=(1+4N)K(\kappa)$. The newly introduced constant $\mu$ needs some
explanation. This is an integration constant coming from the solution of the
Dyson--Schwinger equation for the one-point correlation function, with the
dimension of an energy. It sets the scale for our non-perturbative solution.
Therefore, it should be related to the constant $\Lambda_{\rm QCD}$ that
emerges from dimensional transmutation in perturbative computations, setting
the scale for the confined phase of the Yang--Mills theory. The solution
$\hat\phi_0(kx)$ can be expanded in a Fourier series,
\begin{eqnarray}
\hat\phi_0(kx)&=&\frac{2\eta}{\sqrt\kappa}\sqrt{\frac{-2k^2\kappa}\lambda}
  \sum_{n=0}^\infty\frac{q^{n+1/2}}{1-q^{2n+1}}\sin\left(\frac{2n+1}2
  \eta(kx+\theta)\right)\ =\nonumber\\
  &=&-2i\eta\sqrt{\frac{2k^2}\lambda}\sum_{n=0}^\infty
  \frac{(-1)^nq^{n+1/2}}{1-q^{2n+1}}\cos\left(\frac{2n+1}2\eta kx\right)
  \ =\nonumber\\
  &=&-i\mu\eta\sum_{n=0}^\infty\frac{(-1)^nq^{n+1/2}}{1-q^{2n+1}}
  (e^{(n+1/2)i\eta kx}+e^{-(n+1/2)i\eta kx})\ =\nonumber\\
  &=&-i\mu\eta\sum_{n=-\infty}^\infty b_{2n+1}e^{(n+1/2)i\eta kx}
  \ =\ -i\mu\eta\sum_{m{\rm\,odd}}b_me^{im\eta kx/2},
\end{eqnarray}
with $\eta=\pi/K(\kappa)$. For non-negative values of $n$ one obviously has
\begin{equation}
b_{2n+1}=\frac{(-1)^nq^{n+1/2}}{1-q^{2n+1}}.
\end{equation}
However, this can also be extended to negative values, such as for
\begin{eqnarray}
b_{-2n-1}&=&\frac{(-1)^{n-1}q^{-n-1/2}}{1-q^{-2n-1}}
  \ =\ \frac{-(-1)^nq^{2n+1}q^{-n-1/2}}{q^{2n+1}(1-q^{-2n-1})}\ =\nonumber\\
  &=&\frac{-(-1)^nq^{n+1/2}}{q^{2n+1}-1}\ =\ \frac{(-1)^nq^{n+1/2}}{1-q^{2n+1}}
  \ =\ b_{2n+1}.
\end{eqnarray}
First order non-locality effects are taken into account by solving
\begin{equation}
(\dAlem+\Delta m_G^2)\underline\phi_1(x)=-\lambda e^{\frac12f(\dalem)}
  [e^{\frac12f(\dalem)}\underline\phi_0(x)]e^{-f(\dalem)}[e^{\frac12f(\dalem)}
  \underline\phi_0(x)]^2.
\end{equation}
For the solution $\underline\phi_1(x)=\hat\phi_1(kx)$ we can use a similar
ansatz
\begin{equation}
\hat\phi_1(kx)=-i\eta\mu\sum_{m\,\rm odd}b_me^{im\eta kx/2},
\end{equation}
where the $b_m$ are to be computed. Substituting
$\underline\phi_0(x)=\hat\phi_0(kx)$ on the right hand side and
$\underline\phi_1(x)=\hat\phi_1(kx)$ on the left hand side, on the right hand
side one calculates step by step
\begin{eqnarray}
\lefteqn{[e^{\frac12f(\dalem)}\underline\phi_0(x)]^2
  \ =\ -\eta^2\mu^2\sum_{m_2,m_3\,\rm odd}b_{m_2}b_{m_3}
  e^{\frac12f(-m_2^2\eta^2k^2/4)}e^{\frac12f(-m_3^2\eta^2k^2/4)}
  e^{i(m_2+m_3)\eta kx/2},}\nonumber\\
\lefteqn{e^{-f(\dalem)}[e^{\frac12f(\dalem)}\underline\phi_0(x)]^2
  \ =\ -\eta^2\mu^2\sum_{m_2,m_3\,\rm odd}b_{m_2}b_{m_3}
  e^{-f(-(m_2+m_3)^2\eta^2k^2/4)}\strut}\nonumber\\&&\strut\times
  e^{\frac12f(-m_2^2\eta^2k^2/4)}e^{\frac12f(-m_3^2\eta^2k^2/4)}
  e^{i(m_2+m_3)\eta kx/2},\nonumber\\
\lefteqn{[e^{\frac12f(\dalem)}\underline\phi_0(x)]e^{-f(\dalem)}
  [e^{\frac12f(\dalem)}\underline\phi_0(x)]^2
  \ =\ i\eta^3\mu^3\sum_{m_1,m_2,m_3\,\rm odd}
  b_{m_1}b_{m_2}b_{m_3}\strut}\nonumber\\&&\strut\times
  e^{\frac12f(-m_1^2\eta^2k^2/4)}e^{-f(-(m_2+m_3)^2\eta^2k^2/4)}
  e^{\frac12f(-m_2^2\eta^2k^2/4)}e^{\frac12f(-m_3^2\eta^2k^2/4)}
  e^{i(m_1+m_2+m_3)\eta kx/2},\nonumber\\
\lefteqn{e^{\frac12f(\dalem)}[e^{\frac12f(\dalem)}\underline\phi_0(x)]
  e^{-f(\dalem)}[e^{\frac12f(\dalem)}\underline\phi_0(x)]^2
  \ =\ i\eta^3\mu^3\sum_{m_1,m_2,m_3\,\rm odd}
  b_{m_1}b_{m_2}b_{m_3}e^{\frac12f(-(m_1+m_2+m_3)^2\eta^2k^2/4)}
  \strut}\nonumber\\&&\strut\times
  e^{\frac12f(-m_1^2\eta^2k^2/4)}e^{-f(-(m_2+m_3)^2\eta^2k^2/4)}
  e^{\frac12f(-m_2^2\eta^2k^2/4)}e^{\frac12f(-m_3^2\eta^2k^2/4)}
  e^{i(m_1+m_2+m_3)\eta kx/2}.\kern51pt
\end{eqnarray}
Therefore, the coefficients $b_m$ are determined by the following system of
equations
\begin{eqnarray}
\lefteqn{b_m(-m^2\eta^2k^2/4+\Delta m_G^2)\ =\ \lambda\eta^2\mu^2
  \sum_{m_1,m_2,m_3\,\rm odd}^{m_1+m_2+m_3=m}b_{m_1}b_{m_2}b_{m_3}
  \strut}\\&&\strut\times
  e^{\frac12f(-(m_1+m_2+m_3)^2\eta^2k^2/4)}e^{-f(-(m_2+m_3)^2\eta^2k^2/4)}
  e^{\frac12f(-m_1^2\eta^2k^2/4)}e^{\frac12f(-m_2^2\eta^2k^2/4)}
  e^{\frac12f(-m_3^2\eta^2k^2/4)}.\nonumber
\end{eqnarray}
One can solve for $b_m$ to obtain
\begin{eqnarray}
\lefteqn{b_m\ =\ \lambda\eta^2\mu^2\sum_{m_1,m_2,m_3\,\rm odd}^{m_1+m_2+m_3=m}
  b_{m_1}b_{m_2}b_{m_3}(-(m_1+m_2+m_3)^2\eta^2k^2/4+\Delta m_G^2)^{-1}
  \strut}\\&&\strut\times
  e^{\frac12f(-(m_1+m_2+m_3)^2\eta^2k^2/4)}e^{-f(-(m_2+m_3)^2\eta^2k^2/4)}
  e^{\frac12f(-m_1^2\eta^2k^2/4)}e^{\frac12f(-m_2^2\eta^2k^2/4)}
  e^{\frac12f(-m_3^2\eta^2k^2/4)},\nonumber
\end{eqnarray}
and inserting back to $\underline\phi_1(x)$, as first iteration one has
\begin{eqnarray}
\lefteqn{\underline\phi_1(x)\ =\ i\lambda\eta^3\mu^3
  \sum_{m_1,m_2,m_3\,\rm odd}\frac{b_{m_1}b_{m_2}b_{m_3}
  e^{i(m_1+m_2+m_3)\eta kx/2}}{(m_1+m_2+m_3)^2\eta^2k^2/4-\Delta m_G^2}
  \strut}\\&&\strut\times
  e^{\frac12f(-(m_1+m_2+m_3)^2\eta^2k^2/4)}e^{-f(-(m_2+m_3)^2\eta^2k^2/4)}
  e^{\frac12f(-m_1^2\eta^2k^2/4)}e^{\frac12f(-m_2^2\eta^2k^2/4)}
  e^{\frac12f(-m_3^2\eta^2k^2/4)}.\nonumber
\end{eqnarray}
However, this is not the physical field. Instead, going back to physics we
have to revert the field redefinition, obtaining
$\phi_1(x)=e^{\frac12f(\dalem)}\underline\phi_1(x)$ or
\begin{eqnarray}\label{phi1f}
\lefteqn{\phi_1(x)\ =\ i\lambda\eta^3\mu^3\sum_{m_1,m_2,m_3\,\rm odd}
  \frac{b_{m_1}b_{m_2}b_{m_3}e^{i(m_1+m_2+m_3)\eta kx/2}}{(m_1+m_2
  +m_3)^2\eta^2k^2/4-\Delta m_G^2}\strut}\\&&\strut\times
  e^{f(-(m_1+m_2+m_3)^2\eta^2k^2/4)}e^{-f(-(m_2+m_3)^2\eta^2k^2/4)}
  e^{\frac12f(-m_1^2\eta^2k^2/4)}e^{\frac12f(-m_2^2\eta^2k^2/4)}
  e^{\frac12f(-m_3^2\eta^2k^2/4)}.\nonumber
\end{eqnarray}
Due to the denominator factor, for $f(\dAlem)=\dAlem/M^2$ the series turns out
to be convergent.

The same is to be performed for Eq.~(\ref{G2eq}) in case of $D=4$. Inserting
$\underline\phi(x)=\underline\phi_0(x)$, in momentum space one has
\begin{eqnarray}
\lefteqn{-(p^2-\Delta m_G^2e^{\frac12f(-p^2)})\underline{\tilde G}_2(p)
  -ie^{\frac12f(-p^2)}\ =}\\
  &=&\frac34\lambda\mu^2\eta^2\sum_{m,n{\rm\,odd}}b_mb_n
  e^{\frac12f(-m^2\eta^2k^2/4)}e^{\frac12f(-n^2\eta^2k^2/4)}
  e^{\frac12f(-(p+\frac12(m+n)\eta k)^2)}
  \underline{\tilde G}_2(p+\frac12(m+n)\eta k).\nonumber
\end{eqnarray}
The localised equation has been solved in Appendix~\ref{SecB} and, according
to the K\"all\'en--Lehmann theorem, provides a spectrum of harmonics, known as
glueball states, as can be understood by the form of the glueball operators in
the Yang--Mills theory according to our solution. The Gaussian factors can be
used to restrict the solution to the lowest harmonics $m,n=\pm 1$, resulting
in
\begin{eqnarray}
\lefteqn{-(p^2-\Delta m_G^2e^{\frac12f(-p^2)})\underline{\tilde G}_2(p)
  -ie^{\frac12f(-p^2)}\ =}\nonumber\\
  &=&-m_G^2\left(e^{\frac12f(-p^2)}\underline{\tilde G}_2(p)
  +\frac12e^{\frac12f(-(p-\eta k)^2)}\underline{\tilde G}_2(p-\eta k)
  +\frac12e^{\frac12f(-(p+\eta k)^2)}\underline{\tilde G}_2(p+\eta k)\right),
  \qquad
\end{eqnarray}
where
\begin{equation}
m_G^2=-\frac32\lambda\mu^2\eta^2b_1^2e^{f(-\eta^2k^2/4)}
  =3\kappa\eta^2k^2b_1^2e^{f(-\eta^2k^2/4)}
\end{equation}
The equation can be solved for $\underline{\tilde G}_2(p)$ to result in
\begin{eqnarray}\label{G2it}
\underline{\tilde G}_2(p)=\underline{\tilde\Delta}(p)
  \Bigg[1+\frac i2m_G^2e^{-\frac12f(-p^2)}
  \left(e^{\frac12f(-(p-\eta k)^2)}\underline{\tilde G}_2(p-\eta k)
  +e^{\frac12f(-(p+\eta k)^2)}\underline{\tilde G}_2(p+\eta k)\right)\Bigg],
  \quad
\end{eqnarray}
where
\begin{equation}
\underline{\tilde\Delta}(p):=\frac{-ie^{\frac12f(-p^2)}}{p^2
  -(m_G^2+\Delta m_G^2)e^{\frac12f(-p^2)}}.
\end{equation}
Eq.~(\ref{G2it}) can be solved iteratively up to arbitrary orders of $m_G^2$,
as it is shown in Ref.~\cite{Frasca:2020jbe}. For simplicity, here we take
only the leading order approximation
$\underline{\tilde G}_2(p)\approx\underline{\tilde\Delta}(p)$.

\begin{table}[th]\begin{center}
\caption{\label{Tab1}\it Values from Ref.~\cite{Athenodorou:2021qvs} for the
dynamical glueball mass $M_G^{\rm lat}$ on the lattice in units of $\sqrt{k^2}$
for different values of the number $N_c$ of colours (and corresponding values
of $\beta$), compared to our estimates $M_G^{\rm est}$ from the solution of
the mass gap equation~(\ref{glueball}) in the local limit $M\to\infty$}
\vspace{12pt}
\begin{tabular}{|c|c|c|c|c|}\hline

$N_c$&$\beta$&$M_G^{\rm lat}/\sqrt{k^2}$&$M_G^{\rm est}/\sqrt{k^2}$&error
  \\\hline
$2$&$2.427$&$3.781(23)$&$3.56137$&$-5.8\%$\\
$3$&$6.235$&$3.405(21)$&$3.25530$&$-4.4\%$\\
$4$&$11.02$&$3.271(27)$&$3.26827$&$-0.1\%$\\
$5$&$17.61$&$3.156(31)$&$3.21857$&$+2.0\%$\\
$6$&$25.35$&$3.102(32)$&$3.21930$&$+3.8\%$\\
$8$&$45.50$&$3.099(26)$&$3.19833$&$+3.2\%$\\
$10$&$71.38$&$3.102(37)$&$3.18957$&$+2.8\%$\\
$12$&$103.03$&$3.156(33)$&$3.18443$&$+1.1\%$\\\hline
\end{tabular}
\end{center}\end{table}

\section{Mass gap equation for the glueball\label{SecGB}}
\begin{figure}[hb]\begin{center}
\epsfig{figure=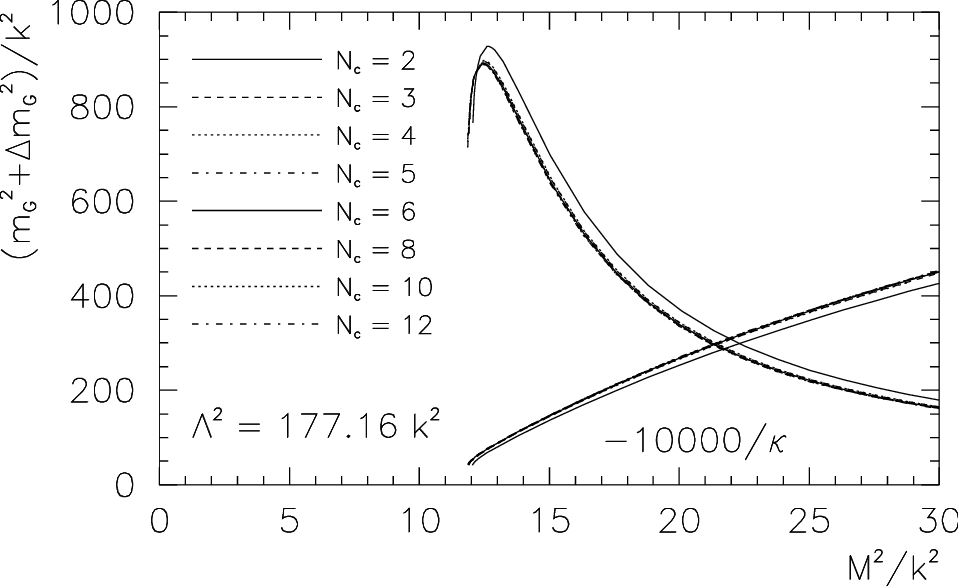, scale=0.8}\quad
\caption{\label{ggap}\it Solution of the mass gap equation~(\ref{glueball})
for different values of $N_c$ in dependence on the non-local scale $M^2$. A
second couple of curves displays the dependence of the dynamically determined
parameter $-10000/\kappa$ on $M^2$, indicating a nontrivial solution of the
gap equation for $\kappa\ne-1$.}
\end{center}\end{figure}

The mass gap equation obtained from
$\Delta m_G^2=-(D-1)\lambda e^{\frac12f(\dalem_x)}G_2(x-x')|_{x'=x}$ reads
\begin{eqnarray}\label{glueball}
\Delta m_G^2&=&\int\frac{d^4p}{(2\pi)^4}\frac{3i\lambda e^{f(-p^2)}}{p^2
  -(m_G^2+\Delta m_G^2)e^{\frac12f(-p^2)}}\ =\ -i\int\frac{d^4p_E}{(2\pi)^4}
  \frac{3i\lambda e^{f(p_E^2)}}{-p_E^2-(m_G^2+\Delta m_G^2)
  e^{\frac12f(p_E^2)}}\ =\nonumber\\
  &=&\frac{-3\lambda}{(4\pi)^2}\int_0^\infty\frac{\rho^3e^{f(\rho^2)}
  d\rho}{\rho^2+(m_G^2+\Delta m_G^2)e^{\frac12f(\rho^2)}}.
\end{eqnarray}
Note that both $m_G^2$ as defined before and $\Delta m_G^2=(1+\kappa)k^2$
contain the common factor $k^2$. This common factor sets the scale of our
solution, as it is proportional to $\mu^2$, the scale of confinement in
Yang--Mills theory as discussed above. Therefore, if we would consider also
quarks, $\mu$ would be related to $\Lambda_{\rm QCD}$ as the scale of
asymptotic freedom. The lattice community will recognise this common factor as
what they call the ``string tension''. At this stage, we do not want to
evaluate the inter-quark potential to see the full relation between all these
three experimental constants. What we need to compare with lattice data are
just the pure numbers arising from the ratio with such general confinement
scales. While the integral is regular at the lower limit, it has to be
regularised for the upper limit. We emphasise that the fact that there is a
non-local factor does not imply at all that all the integrals in the theory
are UV-finite. What such an approach grants is a physical cut-off scale $M$
given by the theory, working like a horison in keeping such integrals finite.
Dimensional regularisation is not applicable in this case, as the integral
cannot be computed analytically, differently from the local limit
$M\to\infty$. One might think of $\rho^2=M^2$ as an upper cut-off. In this
case the integral will diverge as $M^2$, and a truncation of this highest
power is necessary in order to obtain a finite local limit. Instead, our
proposal is to use an upper cut-off $\Lambda^2$ fixed up to the scale $k^2$
which is determined by matching the result to values obtained on the lattice
for different values of $N_c=2,3,4,5,6,8,10,12$~\cite{Athenodorou:2021qvs}.
Results from the older Ref.~\cite{Lucini:2004my} were analysed in the local
case in Refs.~\cite{Frasca:2017mrh,Frasca:2017slg} with excellent agreement.
The idea is to fix the cut-off scale for the mass gap integral using the
results of the local theory (given for $M\to\infty$), in order to prevent
change to the physics with the introduction of the non-locality. Such changes,
if any, do not appear experimentally and they should be really tiny. In order
to compare with the lattice, instead of $\lambda$ as input one has to use
$\lambda=N_cg_s^2=2N_c^2/\beta$ with $\beta=2N_c/g_s^2$. The values for
$\beta$ are taken from Ref.~\cite{Athenodorou:2021qvs} and listed in the
second column of Tab.~\ref{Tab1}.\footnote{A recent paper about the Casimir
effect in non-Abelian gauge theories on the lattice~\cite{Chernodub:2023dok}
has shown that the ground state of the local Yang--Mills theory is not the
same as that found in lattice computations (cf.\ Tab.~\ref{Tab1}). The
authors of Ref.~\cite{Chernodub:2023dok} obtain $M_G^{\rm lat}=1.0(1)$ that
corresponds for us to the choice $n=0$ in the glueball spectrum displayed in
Appendix~B and Refs.~\cite{Frasca:2017mrh,Frasca:2017slg}. This is also in
agreement with the picture of the Casimir effect in local Yang--Mills theory
presented in Ref.~\cite{Frasca:2011fj}.} The agreement is excellent. For the
fit of the mass values in the local limit we use the method of least squares
by minimising
\begin{equation}
\sigma(\Lambda)=\frac1N\sum_{i=1}^N(M_G^{(i)\rm est}(\Lambda)
-M_G^{(i)\rm lat})^2,
\end{equation}
with $M_G^2=m_G^2+\Delta m_G^2$. The result is
given by $\Lambda^2=177.16(3)k^2$. As this result suggests, the agreement
between the lattice values and our estimates is very good, as these are found
in the remaining lines of Tab.~\ref{Tab1}. In Fig.~\ref{ggap} we show the
solution of the mass gap equation~(\ref{glueball}) for $M_G^2$ for the upper
limit $\Lambda^2=177.16k^2$ in dependence on the non-local scale $M^2$,
together with the dependence of the parameter $\kappa$. This should possibly
be of the order of $\TeV^2$. It is interesting to point out that, in this way,
we can obtain a physical understanding of the non-local scale and its proper
order of magnitude as compared to local physics.

\section{Mass gap equation of the quark\label{SecGF}}
\begin{figure}[ht]\begin{center}
\epsfig{figure=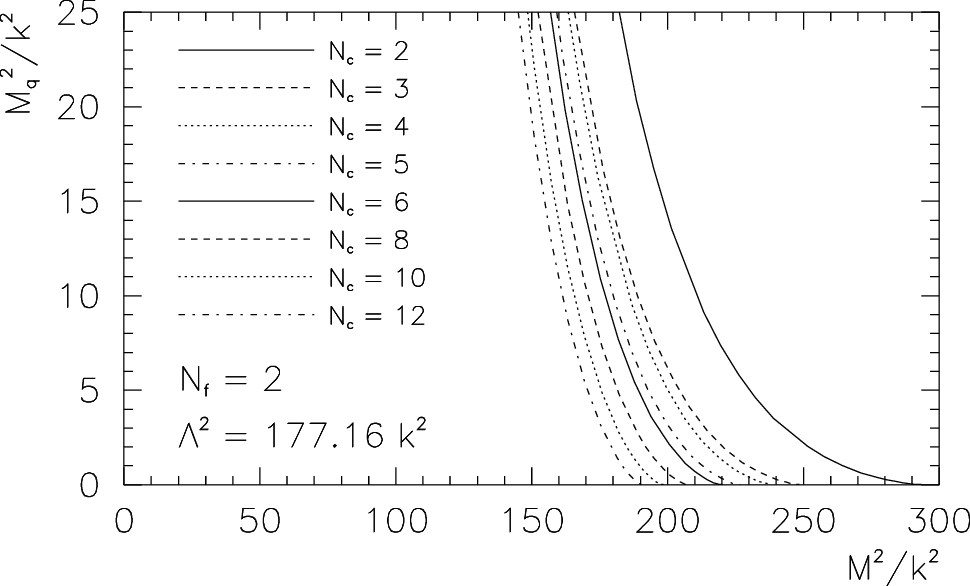, scale=0.8}
\caption{\label{qggap} \it Result of the gap mass determination for the quark
for different values of $N_c$ in dependence on the non-local scale $M^2$. The
number of flavours is set to $N_f=2$.}
\end{center}\end{figure}

The solution obtained in the previous section is input for the determination
of the dynamical quark mass. Namely, the Green function $\underline
G_{2\mu\nu}^{(0)ab}(x,y)=\eta_{\mu\nu}\delta^{ab}\underline G_2(x-y)$ with
$\underline G_2(x-y)$ satisfying the differential equation~(\ref{G2eq}) has
to be convoluted with the left hand side of the equation of motion~(\ref{EL})
in order to obtain
\begin{equation}
\underline A_a^\nu(x)=ig_s\int\underline G_2(x-y)e^{\frac12f(\dalem_y)}
  \bar\psi(y)\gamma^\nu T_a\psi(y)d^4y.
\end{equation}
This result is inserted into the Euler--Lagrange equation for the quark field,
\begin{equation}
0=\frac{\partial{\cal L}}{\partial\bar\psi}-\partial_\mu
  \frac{\partial{\cal L}}{\partial(\partial_\mu\bar\psi)}
  =\frac{\partial{\cal L}}{\partial\bar\psi}=\gamma^\mu
  \left(i\partial_\mu-g_sT_ae^{\frac12f(\dalem)}\underline A_\mu^a\right)\psi
\end{equation}
to obtain
\begin{equation}
0=i\gamma^\mu\partial_\mu\psi(x)+ig_s^2\gamma^\mu T_a\psi(x)
  e^{\frac12f(\dalem)}\int\underline G_2(x-y)e^{\frac12f(\dalem_y)}
  \bar\psi(y)\gamma_\mu T_a\psi(y)d^4y.
\end{equation}
This equation of motion can be understood as the equation of motion of a
non-local Nambu--Jona-Lasinio (NJL) model. After integration by parts, we
derive the NJL Lagrangian
\begin{equation}
{\cal L}_{\rm NJL}=\bar\psi(x)i\gamma^\mu\partial_\mu\psi(x)
  +ig_s^2\bar\psi(x)\gamma^\mu T_a\psi(x)e^{\frac12f(\dalem)}
  \int[e^{\frac12f(\dalem_y)}\underline G_2(x-y)]\bar\psi(y)\gamma_\mu
  T_a\psi(y)d^4y.
\end{equation}
The standard procedure to reach up to the mass gap equation is taken from
Ref.~\cite{Hell:2008cc} and consists of several steps. First of all, a Fierz
rearrangement leads to the action integral
$S[\psi,\bar\psi]=S_0[\psi,\bar\psi]+S_{\rm int}[\psi,\bar\psi]$
with
\begin{eqnarray}
S_0[\psi,\bar\psi]&=&\int d^4x\bar\psi^i(x)i\gamma^\mu\partial_\mu\psi^i(x)
  \quad\mbox{and}\nonumber\\
S_{\rm int}[\psi,\bar\psi]&=&ig_s^2\int\underline G_2^f(x-y)\bar\psi^i(x)
  \Gamma_\alpha^{ij}\psi^j(y)\bar\psi^j(y)\Gamma^\alpha_{ji}\psi^i(x)d^4y
  \ =\nonumber\\
  &=&ig_s^2\int d^4w\int d^4z\bar\psi(w+\frac z2)\Gamma_\alpha\psi(w-\frac z2)
  \underline G_2^f(z)\bar\psi(w-\frac z2)\Gamma^\alpha\psi(w+\frac z2)
\end{eqnarray}
with $\underline G_2^f(z)=e^{f(\dalem)}\underline G_2(z)$. This action
integral is ingredient for the functional integral
\begin{equation}
{\cal Z}=\int{\cal D}\psi{\cal D}\bar\psi\exp(-iS[\psi,\bar\psi]).
\end{equation}
In the following we use $i\underline G_2^f(z)=G\underline C_2(z)/2$ with
$\int\underline C_2(z)d^4z=1$. The quartic interaction term can be removed by
introducing the meson field $(\phi^\alpha)=(\sigma,\vec\pi)$ via the factor
\begin{equation}
\int{\cal D}\phi\exp\left(-\frac{i}{2G}\int d^4z\,\underline C_2(z)
  \int d^4w\phi_\alpha^*(w)\phi^\alpha(w)\right),
\end{equation}
to the functional integral, interchanging the integrations over $w$ and $z$,
and performing the functional ``shift''
\begin{equation}
\phi_\alpha(w)\to\phi_\alpha(w)+g_sG\bar\psi(w-\frac z2)\Gamma_\alpha
  \psi(w+\frac z2).
\end{equation}
In doing so and returning in part to $x$ and $y$, one ends up with the action
functional
\begin{eqnarray}
\lefteqn{S[\psi,\bar\psi,\phi]\ =\ \frac1{2G}\int d^4z\,\underline C_2(z)
  \int d^4w\,\phi_\alpha^*(w)\phi^\alpha(w)+\strut}\nonumber\\&&\strut
  +\int d^4x\int d^4y\,\bar\psi(x)\left[\delta^{(4)}(x-y)i\gamma^\mu
  \partial_\mu-g_s\underline C_2(x-y)\real\phi_\alpha\pfrac{x+y}2
  \Gamma^\alpha\right]\psi(y),\qquad
\end{eqnarray}
The functional integral reads
\begin{equation}
{\cal Z}=\int{\cal D}\psi{\cal D}\bar\psi{\cal D}\phi
  \exp(-iS[\psi,\bar\psi,\phi]).
\end{equation}
Performing a Fourier transform and integrating out the fermionic degrees of
freedom, one ends up with the functional determinant and the bosonised action
\begin{eqnarray}
\lefteqn{S_{\rm bos}[\phi]\ =\ \frac1{2G}\int\frac{d^4q}{(2\pi)^4}
  \phi_\alpha^*(q)\phi^\alpha(q)+\strut}\nonumber\\&&
  +i\ln\det\left[(2\pi)^4\delta^{(4)}(p'-p)\gamma^\mu p_\mu
  -\frac12g_s\underline{\tilde C}_2\pfrac{p'+p}2(\phi_\alpha(p'-p)
  +\phi_\alpha^*(p-p'))\Gamma^\alpha\right],
\end{eqnarray}
where the logarithm of the determinant is understood between momentum states,
$\langle p'|\cdots|p\rangle$. While in the mean field approximation the first
term gives $\bar\sigma^2/2G$, the determinant (for $N_f=2$ flavours) leads to
\begin{equation}
\det(\gamma^\mu p_\mu-M_q(p)\bbbone)=(p^2-M_q^2(p))^{4N_c},
\end{equation}
with $M_q(p)=g_s\underline{\tilde C}_2(p)\bar\sigma$. The variation of the
action integral with respect to the mean value $\bar\sigma$ leads to
\begin{equation}
\bar\sigma=-8iN_cg_sG\int\frac{d^4p}{(2\pi)^4}
  \frac{\underline{\tilde C}_2(p)M_q(p)}{p^2-M_q^2(p)},
\end{equation}
and the insertion into the definition of $M_q(p)$ finally leads to the mass
gap equation
\begin{eqnarray}\label{Qgap}
M_q(p)&=&-8iN_cg_s^2G\underline{\tilde C}_2(p)\int\frac{d^4q}{(2\pi)^4}
  \frac{\underline{\tilde C}_2(q)M_q(q)}{q^2-M_q^2(q)}\ =\nonumber\\
  &=&16\lambda\frac{\tilde G_2(p)}{\underline{\tilde G}_2(0)}\int
  \frac{d^4q}{(2\pi)^4}\frac{\underline{\tilde G}_2(q)M_q(q)}{q^2-M_q^2(q)},
\end{eqnarray}
where we inserted $\underline{\tilde C}_2(p)=-2i\underline{\tilde G}_2(p)/G$
and used
\begin{equation}
\underline G_2^f(z)=\int\frac{d^4q}{(2\pi)^4}e^{f(\dalem)}
  \underline{\tilde G}_2(q)e^{-iqz}
\end{equation}
to obtain
\begin{equation}
\frac G2=i\int\underline G_2^f(z)d^4z
  =i\int\int\frac{d^4q}{(2\pi)^4}e^{f(-q^2)}
  \underline{\tilde G}_2(q)e^{-iqz}d^4z=i\underline{\tilde G}_2(0).
\end{equation}
In order to solve the gap equation for the quark, we assume that the gap mass
$M_q(p)$ does not depend explicitly on the momentum. In this case one has to
solve the equation
\begin{equation}
M_q=16\lambda\int\frac{d^4q}{(2\pi)^4}\frac{e^{f(-q^2)}
  \underline{\tilde G}_2(q)M_q}{q^2-M_q^2}.
\end{equation}
In principle, $M_q$ can be cancelled on both sides, leading to a gap equation
similar to Eq.~(3.7) in the original NJL publication~\cite{Nambu:1961tp}.
However, in order to solve this equation iteratively, it is more appropriate
to instead multiply it with $M_q$ in order to determine $M_q^2$ from 
\begin{eqnarray}
M_q^2&=&16N_cg_s^2\int\frac{d^4q}{(2\pi)^4}
  \frac{e^{f(-q^2)}\underline{\tilde G}_2(q)M_q^2}{q^2-M_q^2}
  \ =\ \int\frac{d^4q}{(2\pi)^4}\frac{-16i\lambda
  M_q^2e^{\frac32f(-q^2)}}{(q^2-M_G^2e^{\frac12f(-q^2)})(q^2-M_q^2)}
  \ =\nonumber\\
  &=&\int\frac{d^4q_E}{(2\pi)^4}\frac{16\lambda M_q^2e^{\frac32f(q_E^2)}}
  {(-q_E^2-M_G^2e^{\frac12f(q_E^2)})(-q_E^2-M_q^2)}\ =\nonumber\\
  &=&\frac{8\lambda}{(4\pi)^2}\int_0^{\Lambda^2}
  \frac{M_q^2\rho^2e^{\frac32f(\rho^2)}d\rho^2}{(\rho^2
  +M_G^2e^{\frac12f(\rho^2)})(\rho^2+M_q^2)}.
\end{eqnarray}
The upper limit is taken to be the same as in the previous section, and the
solution $M_G^2$ of the mass gap equation for the glueball is used. The fixed
point problem converges, and one obtains a dependence on the non-local scale
$M^2$ which is displayed in Fig.~\ref{qggap}. We can see that chiral symmetry
breaking appears in non-local QCD as well.

\begin{figure}[ht]\begin{center}
\epsfig{figure=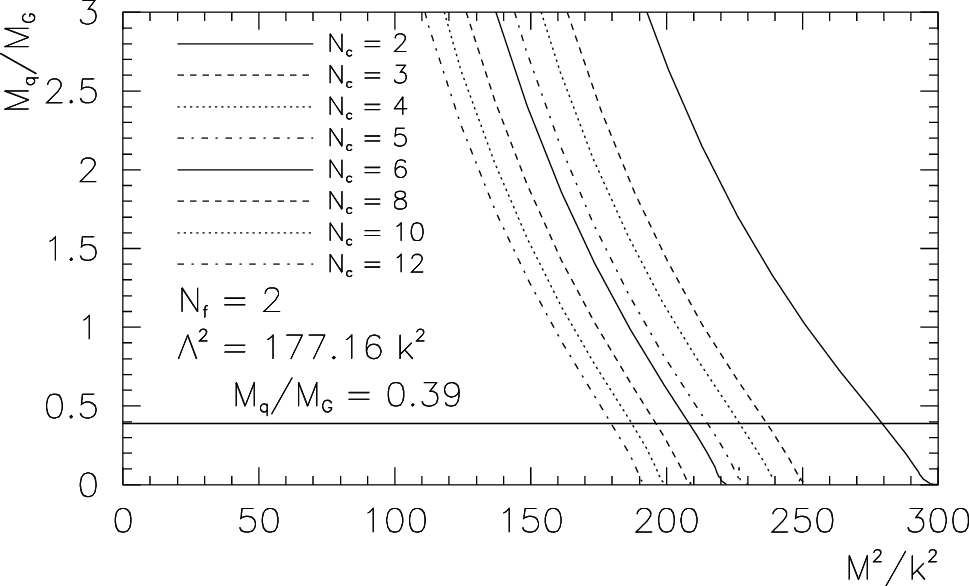,scale=0.8}
\caption{\label{qgrap} \it Quark--glueball mass ratio $M_q/M_G$ in dependence
on $M^2$, as read off from Figs.~\ref{ggap} and~\ref{qggap}. Shown is the
threshold value $M_q/M_G=0.39$.}
\end{center}\end{figure}

It is shown in Refs.~\cite{Frasca:2022lwp,Frasca:2022pjf} that an educated
guess for confinement is given by the threshold $M_q/M_G>0.39$, above which
two real pole solutions for $p^2$ in the equation
\begin{equation}
p^2=M_q^2\pfrac{\tilde G_2(p)}{\tilde G_2(0)}^2
  =\frac{M_q^2M_G^4}{(p^2-M_G^2)^2}
\end{equation}
change to imaginary pole solutions. According to Fig.~\ref{qgrap}, this is
the case for the non-local scale $M^2$ below $150k^2$.

\section{Discussion and Conclusions\label{SecC}}

The spectrum of the Yang--Mills field is analysed for the non-local
Yang--Mills sector and found in agreement with the local results on lattice in
the limit of the non-local scale running to infinity (cf.\ Tab.~\ref{Tab1}).
The agreement with lattice is astonishingly good, and also with respect to
previous computations~\cite{Frasca:2017mrh,Frasca:2017slg}. Because of this,
this spectrum becomes our reference for the non-local case. We point out that
recent lattice results about the Casimir effect in Yang--Mills theory seem to
challenge the determination of the ground state of the theory on lattice. In
our case, this would be a further confirmation of our approach (for details
see Ref.~\cite{Chernodub:2023dok}). It is a relevant result of our
investigations that scale invariance is badly broken by interactions, a fact
that can be taken as a possible clue of confinement. Indeed, the solution of
the gap equation for the fermion shows some indication of quark confinement
also in the non-local case. This result is really important, as it seems to
point to the fact that confinement could be an ubiquitous effect in nature
that removes degrees of freedom in a theory to favour others. A rigorous
mathematical proof of confinement is beyond the scope of the present
manuscript. In general, further studies are needed to improve these results.
Still, the results obtained here appear as a sound confirmation of previous
work with a different technique and with the important extension to non-local
QCD.

\section*{Acknowledgements}
The research was supported in part by the European Regional Development Fund
under Grant No.~TK133.

\begin{appendix}

\section{Dyson--Schwinger equations\label{SecA}}
In this appendix we show how to derive the Dyson--Schwinger equations for a
$\phi^4$ theory in differential form, as exemplified in
Refs.~\cite{Frasca:2013tma,Frasca:2015yva} using the technique devised in
Ref.~\cite{Bender:1999ek} by Bender, Milton and Savage. Such a technique was
initially conceived for a ${\cal PT}$-invariant non-Hermitian theory and
properly extended to more general cases by one of us (M.F.). The starting
point is the (massless) Lagrangian
\begin{equation}
{\cal L}=\frac12(\partial\phi)^2-\frac14\lambda\phi^4,
\end{equation}
leading to the generating functional
\begin{equation}
Z[j]=\int[d\phi]\exp\left(i\int({\cal L}+j\phi)d^4x\right).
\end{equation}
First of all, after integration by parts the variation of this generating
functional with respect to $\phi$ leads to the Dyson--Schwinger master
equation
\begin{equation}
\frac1{Z[j]}\frac{-i\delta Z[j]}{\delta\phi(x)}
  =-\langle\partial^2\phi(x)\rangle_j-\lambda\langle\phi^3(x)\rangle_j+j(x)=0,
\end{equation}
where
\begin{equation}
\langle O(x)\rangle_j=\frac{\int[d\phi]O(x)\exp\left(i\int({\cal L}+j\phi)d^4x
  \right)}{\int[d\phi]\exp\left(i\int({\cal L}+j\phi)d^4x\right)}\Big|_{j=0}
  =\frac{\int[d\phi]O(x)\exp\left(i\int{\cal L}d^4x\right)}{\int[d\phi]
  \exp\left(i\int{\cal L}d^4x\right)}.
\end{equation}
Following a notation used e.g.\ by Abbott~\cite{Abbott:1981ke}, the
application of functional derivatives with respect to the current $j$ to the
effective action $W[j]=\ln Z[j]$ on the one and to the generating
functional $Z[j]$ on the other hand leads to the tower of Green functions
\begin{eqnarray}\label{Green}
G_1^j(x)&=&\pfrac{-i\delta}{\delta j(x)}W[j]\ =\ \frac1{Z[j]}
  \pfrac{-i\delta}{\delta j(x)}Z[j]\ =\ \langle\phi(x)\rangle_j,\nonumber\\
G_2^j(x,y)&=&\pfrac{-i\delta}{\delta j(x)}\pfrac{-i\delta}{\delta j(y)}W[j]
  \ =\ \pfrac{-i\delta}{\delta j(x)}\frac1{Z[j]}\pfrac{-i\delta}{\delta j(y)}
  Z[j]\ =\nonumber\\
  &=&-\frac1{Z[j]^2}\left(\frac{-i\delta}{\delta j(x)}Z[j]\right)\left(
  \frac{-i\delta}{\delta j(y)}Z[j]\right)+\frac1{Z[j]}
  \pfrac{-i\delta}{\delta j(x)}\pfrac{-i\delta}{\delta j(y)}Z[j]
  \ =\nonumber\\[7pt]
  &=&-\langle\phi(x)\rangle_j\langle\phi(y)\rangle_j
  +\langle\phi(x)\phi(y)\rangle_j,\nonumber\\[7pt]
G_3^j(x,y,z)
  &=&2\langle\phi(x)\rangle_j\langle\phi(y)\rangle_j\langle\phi(z)\rangle_j
  -\langle\phi(x)\rangle_j\langle\phi(y)\phi(z)\rangle_j
  -\langle\phi(y)\rangle_j\langle\phi(z)\phi(x)\rangle_j
  +\strut\nonumber\\[3pt]&&\strut
  -\langle\phi(z)\rangle_j\langle\phi(x)\phi(y)\rangle_j
  +\langle\phi(x)\phi(y)\phi(z)\rangle_j,\ \ldots
\end{eqnarray}
Inverting this system step by step one obtains the tower
\begin{eqnarray}
\langle\phi(x)\rangle_j&=&G_1^j(x),\nonumber\\[3pt]
\langle\phi(x)\phi(y)\rangle_j&=&G_2^j(x,y)+G_1^j(x)G_1^j(y),\nonumber\\[3pt]
\langle\phi(x)\phi(y)\phi(z)\rangle_j&=&G_3^j(x,y,z)+G_2^j(x,y)G_1^j(z)
  +G_2^j(y,z)G_1^j(x)+G_2^j(z,x)G_1^j(y)+\strut\nonumber\\[3pt]&&\strut
  +G_1^j(x)G_1^j(y)G_1^j(z),\ \ldots
\end{eqnarray}
\begin{eqnarray}
\raise-48pt\hbox{\epsfig{figure=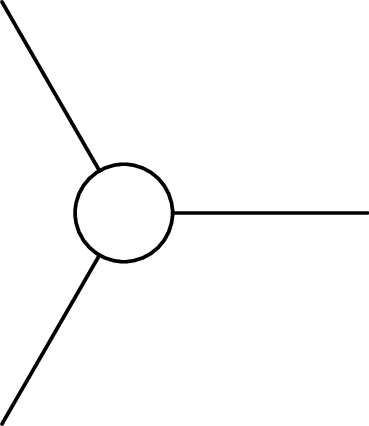, scale=0.5}}
  &=&\raise-48pt\hbox{\epsfig{figure=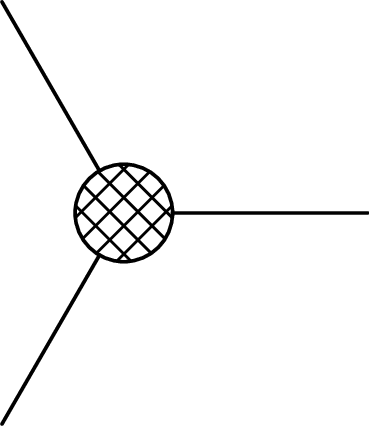, scale=0.5}}\ +\
  3\quad\raise-48pt\hbox{\epsfig{figure=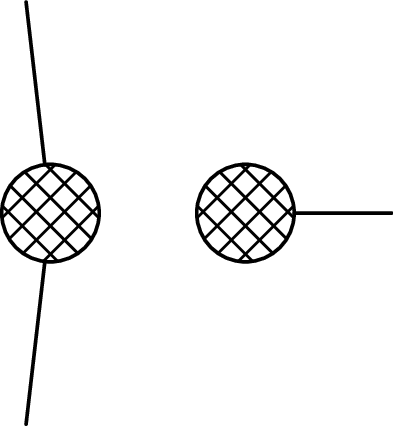, scale=0.5}}\ +\
  \raise-48pt\hbox{\epsfig{figure=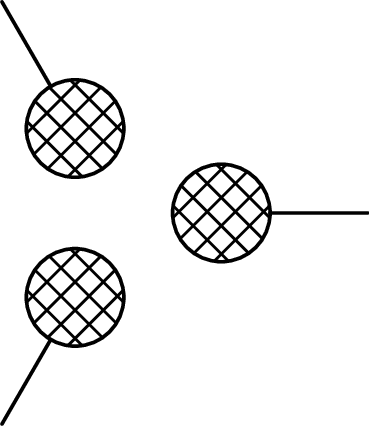, scale=0.5}}\nonumber
\end{eqnarray}
This tower is finished by the expectation value of the Dyson--Schwinger master
equation which via insertion of the tower leads to
\begin{eqnarray}\label{eqj}
j(x)&=&\partial^2\langle\phi(x)\rangle_j+\lambda\langle\phi^3(x)\rangle_j
  \ =\nonumber\\[3pt]
  &=&\partial^2G_1^j(x)+\lambda\left(G_3^j(x,x,x)+3G_2^j(x,x)G_1^j(x)
  +G_1^j(x)^3\right).
\end{eqnarray}

In obtaining equations for the Green functions by setting $j=0$, at the same
time we use translation invariance to write the Green functions in the form
\begin{equation}
G_1(x):=G_1^j(x),\qquad
G_k(x_1-x_2,x_2-x_3,\ldots,x_{k-1}-x_k):=G_k^j(x_1,x_2,x_3,\ldots,x_{k-1},x_k)
\end{equation}
to obtain
\begin{equation}\label{eqG1}
\partial^2G_1(x)+\lambda\left(G_3(0,0)+3G_2(0)G_1(x)+G_1(x)^3\right)=0.
\end{equation}
As $G_2(0)$ and $G_3(0,0)$ are constants, this equation is an ordinary
(though non-linear) differential equation for $G_1(x)$. For
$3\lambda G_2(0)=\Delta m_G^2$ and $G_3(0,0)=0$ one has
$\partial^2G_1+\Delta m_G^2G_1+\lambda G_1^3=0$. As shown in
Ref.~\cite{Frasca:2015yva}, this non-linear differential equation is solved by
\begin{equation}\label{solG1}
G_1(x)=\mu\sn(k\cdot x+\theta|\kappa),\qquad
  \mu=\sqrt{\frac{2(k^2-\Delta m_G^2)}\lambda}
\end{equation}
with
\begin{equation}
k^2=\Delta m_G^2+\frac12\lambda\mu^2,\qquad
\kappa=\frac{\Delta m_G^2-k^2}{k^2},
\end{equation}
where $\mu$ and $\theta$ are integration constants, and $\sn(\zeta|\kappa)$,
$\cn(\zeta|\kappa)$ and $\dn(\zeta|\kappa)$ are Jacobi elliptic functions.
Even for $\Delta m_G=0$, i.e.\ for the absence of a Green function or gap
mass, one obtains a nontrivial solution
\begin{equation}\label{solG10}
G_1(x)=\sqrt{\frac{2k^2}\lambda}\sn(k\cdot x+\theta|-1),\qquad
  k^2=\frac12\lambda\mu^2.
\end{equation}

If we take the functional derivative $-i\delta/\delta j(y)$ of Eq.~(\ref{eqj})
before setting $j=0$ one obtains the differential equation
\begin{eqnarray}
\lefteqn{-i\delta^4(x-y)=\partial^2G_2^j(x,y)+\strut}\nonumber\\[3pt]&&\strut
  +\lambda\left(G_4^j(x,x,x,y)+3G_3^j(x,x,y)G_1^j(x)
  +3G_2^j(x,x)G_2^j(x,y)+3G_1^j(x)^2G_2^j(x,y)\right).\qquad
\end{eqnarray}
Again, we can choose $G_4^j(x,x,x,y)=G_3^j(x,x,y)=0$ and
$3\lambda G_2^j(x,x)=\Delta m_G^2$ to obtain
\begin{equation}
\partial^2G_2^j(x,x')+\Delta m_G^2G_2^j(x,x')+3\lambda G_1(x)^2G_2^j(x,x')
  =-i\delta^4(x-x').
\end{equation}
Note that the Green function defined by this differential equation is not
translational invariant. Therefore, we cannot use $G_2^j(x,x')=G_2(x-x')$ at
this point. However, as shown in Appendix~\ref{SecB} we can restore the
translational invariance. Inserting Eq.~(\ref{solG1}) one obtains
\begin{equation}
\partial^2G_2^j(x,x')+\Delta m_G^2G_2^j(x,x')-6\kappa k^2
  \sn^2(k\cdot x+\theta\,|\kappa)G_2^j(x,x')=-i\delta^4(x-x')
\end{equation}
(note that $\lambda\mu^2=2(k^2-\Delta m_G^2)=-2\kappa k^2$).

In order to show that our solution for $G_1$ is meaningful, we show that the
theory has a zero mode. The Hamiltonian of the system is given by
\begin{equation}
H=\int d^3x\left[\frac12\pi^2(x)+\frac12\left(\nabla\phi(x)\right)^2
  +\frac\lambda4\phi^4(x)\right].
\end{equation} 
We expand this Hamiltonian around the classical solution, given by $G_1(x)$
in Eq.~(\ref{solG1}),
\begin{equation}
\phi(x)=G_1(x)+\delta\phi(x),\qquad
\pi(x)=\partial_tG_1(x)+\partial_t\delta\phi(x),  
\end{equation}
yielding
\begin{equation}
H=H_0+\int d^3x\left[\frac12\left(\partial_t\delta\phi(x)\right)^2
  +\frac12\left(\nabla\delta\phi(x)\right)^2
  +\frac32\lambda G_1^2(x)\delta\phi^2(x)\right]
  +O\left(\delta\phi^3\right),
\end{equation}
where $H_0$ is the contribution coming from the classical solution. The linear
part vanishes after integration by parts and the application of the equations
of motion for the classical solution. The quadratic part can be diagonalised
with a Fourier series provided that we are able to obtain the eigenvalues and
the eigenvectors of the operator
\begin{equation}
L_{\mu_0^2=0}=-\Box+3\lambda G_1^2(x).
\end{equation}
It is not difficult to realise that there is a zero mode. We give the
solutions for both the zero and non-zero modes. The spectrum is continuous
with eigenvalues $0$ and $3\mu^2\lambda/2$, where $\mu$ varies continuously
from zero to infinity. The zero-mode solution has the form
\begin{equation}
\chi_0(x,\mu)=a_0\cn(k\cdot x+\theta|-1)\dn(k\cdot x+\theta|-1),
\end{equation}
where $a_0$ is a normalisation constant. Non-zero modes are given by
\begin{equation}
\chi(x,\mu)=a'\sn(k\cdot x+\theta|-1)\dn(k\cdot x+\theta|-1),
\end{equation}
with $a'$ again a normalisation constant. These equations hold on-shell, that
is when $k^2=\mu^2\lambda/2$. Since the spectrum is continuous, the
eigenfunctions are not normalisable. Therefore, we note that there is a doubly
degenerate set of zero modes spontaneously breaking translational invariance
and the $Z_2$ symmetry of the theory. For the zero mode this results in
\begin{equation}
\chi_0(x,\mu)=-2a_0\frac{\sn(k\cdot x|-1)}{\dn^2(k\cdot x|-1)}.
\end{equation}
For a given parameter $\mu$, $Z_2$ symmetry is spontaneously broken through
this zero mode. The mode disappears when $\mu=0$, as it should, and one gets
back to a standard textbook solution.

\section{2P-correlation function for the scalar field\label{SecB}}
Introducing a parameter $\eps$, the differential equation
\begin{equation}
\partial_t^2G_2^j(x,x')+\Delta m_G^2G_2^j(x,x')-6\kappa k^2
  \sn^2(k\cdot x+\theta\,|\kappa)G_2^j(x,x')
  =-i\delta^4(x-x')+\eps\nabla^2G_2^j(x,x')
\end{equation}
is iteratively solved for $G_2^j(x,x')$ by using the gradient expansion in
$\eps$,
\begin{equation}
G_2^j(x,x')=\sum_{n=0}^\infty\eps^nG_2^{(n)}(x,x')
\end{equation}
where $\eps=1$ is set in the end. One obtains
\begin{eqnarray}\label{gradsys}
\partial_t^2G_2^{(0)}(x,x')+\Delta m_G^2G_2^{(0)}(x,x')
  -6\kappa k^2\sn^2(k\cdot x+\theta\,|\kappa)G_2^{(0)}(x,x')
  &=&-i\delta^4(x-x'),\nonumber\\
\partial_t^2G_2^{(1)}(x,x')+\Delta m_G^2G_2^{(1)}(x,x')
  -6\kappa k^2\sn^2(k\cdot x+\theta\,|\kappa)G_2^{(1)}(x,x')
  &=&\nabla^2G_2^{(0)}(x,x'),\nonumber\\
\partial_t^2G_2^{(2)}(x,x')+\Delta m_G^2G_2^{(2)}(x,x')
  -6\kappa k^2\sn^2(k\cdot x+\theta\,|\kappa)G_2^{(2)}(x,x')
  &=&\nabla^2G_2^{(1)}(x,x'),\nonumber\\&&\ldots
\end{eqnarray}
In order to perform the calculation, we start with the rest frame of the
motion. As the first equation contains no spatial derivative, one has
$G_2^{(0)}(x,x')=\delta^3(\vec x-\vec x\,')\bar G_2(t,t')$ which simplifies
the first differential equation (for the choice $\vec x=\vec 0$ and,
accordingly, $\vec x\,'=\vec 0$) to
\begin{equation}
\bar G''_2(t,t')+\Delta m_G^2\bar G_2(t,t')-6\kappa k^2
  \sn^2(\omega t+\theta\,|\kappa)\bar G_2(t,t')=-i\delta(t-t').
\end{equation}
The corresponding homogeneous differential equation is solved by
$\bar G_2(t):=\bar G_2(t,t')=\hat G_2(\omega t+\theta)$ which does not depend
on $t'$ at all. Therefore, we can set $t'=0$ in the following. The
differential equation for the Green function is then solved by
$\bar G_2(t)=C\Theta(t)\hat G_2(\omega t+\theta)$ with the Heaviside step
function $\Theta(t)$, where the amplitude $C$ and the phase $\theta$ get fixed
as well, and this also holds if the mass term is not skipped. This can be seen
by calculating the derivatives to obtain
\begin{eqnarray}
\bar G_2(t)&=&C\Theta(t)\hat G_2(\omega t+\theta),\nonumber\\[7pt]
\partial_t\bar G_2(t)&=&C\delta(t)\hat G_2(\omega t+\theta)
  +C\Theta(t)\omega\hat G'_2(\omega t+\theta),\nonumber\\[7pt]
\partial_t^2\bar G_2(t)&=&C\delta'(t)\hat G_2(\omega t+\theta)
  +2C\delta(t)\omega\hat G'_2(\omega t+\theta)+C\Theta(t)\omega^2
  \hat G''_2(\omega t+\theta)\ =\nonumber\\[3pt]
  &=&C\delta(t)\omega\hat G'_2(\omega t+\theta)+C\Theta(t)\omega^2
  \hat G''_2(\omega t+\theta)
\end{eqnarray}
(where we used $\delta'(t)f(t)=-\delta(t)f'(t)$) and inserting
\begin{eqnarray}\label{difft}
\lefteqn{C\delta(t)\omega\hat G'_2(\omega t+\theta)+\strut}\\&&\strut\kern-23pt
  +C\Theta(t)\left(\omega^2\hat G''_2(\omega t+\theta)
  +\Delta m_G^2\hat G_2(\omega t+\theta)-6\kappa k^2
  \sn^2(\omega t+\theta\,|\kappa)\hat G_2(\omega t+\theta)\right)=-i\delta(t).
  \nonumber
\end{eqnarray}
For $\Delta m_G=0$ the homogeneous equation is
solved by $\hat G_2(\zeta)=\hat G'_1(\zeta)$ with
$\hat G_1(\zeta)=\sn(\zeta|-1)$. However, using the more general ansatz
$\hat G_1(\zeta)=\sn(\zeta|\kappa)$, this also holds for $\Delta m_G\ne 0$.
Namely, one obtains $\hat G_2(\zeta)=\dn(\zeta|\kappa)\cn(\zeta|\kappa)$,
\begin{eqnarray}
\hat G'_2(\zeta)&=&-\sn(\zeta|\kappa)
  \left(\dn^2(\zeta|\kappa)+\kappa\cn^2(\zeta|\kappa)\right),\nonumber\\
\hat G''_2(\zeta)&=&-\dn(\zeta|\kappa)\cn(\zeta|\kappa)
  \left(\dn^2(\zeta|\kappa)+\kappa\cn^2(\zeta|\kappa)
  -4\kappa\sn^2(\zeta|\kappa)\right).
\end{eqnarray}
Therefore, with $\zeta=\omega t+\theta$ and $\omega^2=k^2$ for the rest frame
one obtains
\begin{eqnarray}
\lefteqn{\omega^2\hat G''_2(\zeta)+\Delta m_G^2\hat G_2(\zeta)
  -6\kappa k^2\sn^2(\zeta|\kappa)\hat G_2(\zeta)\ =}\nonumber\\[3pt]
  &=&\left(-k^2\left(\dn^2(\zeta|\kappa)+\kappa\cn^2(\zeta|\kappa)
  -4\kappa\sn^2(\zeta|\kappa)\right)+\Delta m_G^2-6\kappa k^2
  \sn^2(\zeta|\kappa)\right)\hat G_2(\zeta)\ =\nonumber\\
  &=&\left(-k^2(1+\kappa)+\Delta m_G^2\right)\hat G_2(\zeta)=0,
\end{eqnarray}
where we have used $\cn^2(\zeta|\kappa)+\sn^2(\zeta|\kappa)=1$ and
$\dn^2(\zeta|\kappa)+\kappa\sn^2(\zeta|\kappa)=1$. One can start with
$\sn(\theta|\kappa)=1$ which is satisfied if $\theta=(1+4N)K(\kappa)$, and
as a consequence of this $C\omega(1-\kappa)=i$ which is solved by
$C=i/\omega(1-\kappa)$. Therefore, we end up with
\begin{equation}
\bar G_2(t)=\frac{i\Theta(t)}{\omega(1-\kappa)}
  \sn'(\omega t+\theta\,|\kappa),\qquad\theta=(1+4N)K(\kappa).
\end{equation}

As Jacobi's elliptic functions $\sn(\zeta|\kappa)$, $\cn(\zeta|\kappa)$ and
$\dn(\zeta|\kappa)$ are periodic functions, it is possible to expand them in a
Fourier series. Using the nome $q=\exp(-\pi K^*(\kappa)/K(\kappa))$ where
$K^*(\kappa)=K(1-\kappa)$ and
\begin{equation}
K(\kappa)=\int_0^{\pi/2}\frac{d\theta}{\sqrt{1-\kappa\sin^2\theta}}
  =F(\pi/2|\kappa),
\end{equation}
one has
\begin{equation}
\sn(\zeta|\kappa)=\frac{2\pi}{K(\kappa)\sqrt\kappa}\sum_{n=0}^\infty
  \frac{q^{n+1/2}}{1-q^{2n+1}}\sin\left((2n+1)\frac{\pi\zeta}{2K(\kappa)}
  \right).
\end{equation}
Inserting this Fourier series into our solution for $\bar G_2(t)$ leads to
\begin{equation}
\bar G_2(t)=\frac{i\Theta(t)\pi^2}{\omega(1-\kappa)K(\kappa)^2\sqrt\kappa}
  \sum_{n=0}^\infty(2n+1)\frac{q^{n+1/2}}{1-q^{2n+1}}
  \cos\left((2n+1)\frac{\pi(\omega t+\theta)}{2K(\kappa)}\right)
\end{equation}
with $\theta=(1+4N)K(\kappa)$. Inserting these values for $\theta$, one has
\begin{equation}
\cos\left((2n+1)\left(\frac{\pi\omega t}{2K(\kappa)}
  +\frac\pi2+2\pi N\right)\right)
  =-(-1)^n\sin\left((2n+1)\frac{\pi\omega t}{2K(\kappa)}\right).
\end{equation}
Our result for the Fourier series of the zeroth order Green function is
therefore given by
\begin{equation}
\bar G_2(t)=\frac{-i\Theta(t)\pi^2}{\omega(1-\kappa)K(\kappa)^2\sqrt\kappa}
  \sum_{n=0}^\infty(2n+1)\frac{(-1)^nq^{n+1/2}}{1-q^{2n+1}}
  \sin\left((2n+1)\frac{\pi\omega t}{2K(\kappa)}\right).
\end{equation}
We can define mass states $m_n=(2n+1)m_0$ with $m_0=\pi\omega/(2K(\kappa))$,
forming the spectrum of the gluonic part of the QCD Lagrangian, i.e.\ the
spectrum of glueballs. The Fourier transform leads to terms of the kind
\begin{eqnarray}
\lefteqn{\int_{-\infty}^\infty\Theta(t)\sin(m_nt)e^{-iEt}dt
  \ =\ \int_0^{\infty}\frac1{2i}
  \left(e^{im_nt}-e^{-im_nt}\right)e^{-iEt}dt\ =}\nonumber\\
  &=&\frac1{2i}\left[\frac{e^{-i(E-m_n)t}}{-i(E-m_n)}
  -\frac{e^{i(E+m_n)t}}{-i(E+m_n)}\right]_{t=0}^\infty
  \ =\ -\frac12\left(\frac1{E-m_n}-\frac1{E+m_n}\right)
  \ =\ \frac{-m_n}{E^2-m_n^2}.\qquad\
\end{eqnarray}
Therefore, in using the Feynman propagator convention, we end up with
\begin{equation}\label{G0E0}
\widetilde{\bar G}_2(E)=i\sum_{n=0}^\infty
  \frac{B_n(-1+\Delta m_G^2/k^2)}{E^2-m_n^2+i\epsilon},\quad
  B_n(\kappa):=\frac{(2n+1)^2\pi^3}{2(1-\kappa)K(\kappa)^3\sqrt\kappa}\
  \frac{(-1)^nq^{n+1/2}}{1-q^{2n+1}}.
\end{equation}
In the massless case, i.e.\ $\kappa=-1$, the situation is simplied, as
$K(-1)=1.31103\ldots$ and $K^*(-1)=K(2)=1.31103\ldots(1-i)$. Therefore,
$q=e^{-\pi(1-i)}=-e^{-\pi}$. For the Fourier coefficient one obtains
\begin{equation}
\frac{e^{-(n+1/2)\pi(1-i)}}{1-e^{-(2n+1)\pi(1-i)}}
  =i(-1)^n\frac{e^{-(n+1/2)\pi}}{1+e^{-(2n+1)\pi}},
\end{equation}
and the factor $i$ cancels against $\sqrt\kappa=\sqrt{-1}=i$ in the
denominator. Therefore, one ends up with
\begin{equation}
  B_n(-1):=\frac{(2n+1)^2\pi^3}{4K(-1)^3}\
  \frac{e^{(2n+1)\pi/2}}{1+e^{(2n+1)\pi}}.
\end{equation}
{\it Vice versa\/}, one has
\begin{equation}
\bar G_2(t-t')=\int\frac{dp_0}{2\pi}\widetilde{\bar G}_2(p_0)e^{ip_0(t-t')}
\end{equation}
and
\begin{equation}
G_2^{(0)}(x,x')=\delta^{(3)}(x-x')\int_{-\infty}^\infty\frac{dp_0}{2\pi}
  \widetilde{\bar G}_2(p_0)e^{ip_0(t-t')}=\int\frac{d^4p}{(2\pi)^4}
  \widetilde{\bar G}_2(p_0)e^{ip(x-x')}.
\end{equation}

Having $G_2^{(0)}(x,x')$ at hand, the system~(\ref{gradsys}) can be solved
iteratively by
\begin{eqnarray}
\lefteqn{G_2^{(1)}(x,x')
  \ =\ i\int G_2^{(0)}(x,x_1)\nabla^2G_2^{(0)}(x_1,x')d^4x_1,}\nonumber\\
\lefteqn{G_2^{(2)}(x,x')
  \ =\ i\int G_2^{(0)}(x,x_2)\nabla^2G_2^{(1)}(x_2,x')d^4x_2\ =}\nonumber\\
  &=&-\int\int G_2^{(0)}(x,x_2)\nabla^2G_2^{(0)}(x_2,x_1)\nabla^2
  G_2^{(0)}(x_1,x')d^4x_1d^4x_2,\quad\ldots\qquad
\end{eqnarray}
and $G_2(x,x')$ the sum of all these (for $\eps=1$). In momentum
space, this is just a Dyson series, where the propagators are given by
Eq.~(\ref{G0E0}), and the vertices can be derived from
\begin{equation}
\nabla^2\delta(\vec x-\vec x\,')=\nabla^2\int\frac{d^3p}{(2\pi)^3}
  e^{i\vec p(\vec x-\vec x\,')}=\int\frac{d^3p}{(2\pi)^3}(-\vec p\,^2)
  e^{i\vec p(\vec x-\vec x\,')}
\end{equation}
to be $-\vec p\,^2$ in momentum space. The Dyson series can be resummed
to~\cite{Frasca:2013tma}
\begin{eqnarray}
\tilde G_2(E;\vec p\,)&=&\sum_{n=0}^\infty\frac{iB_n}{E^2-m_n^2+i\epsilon}
  +\sum_{n=0}^\infty\frac{iB_n}{E^2-m_n^2+i\epsilon}(-i\vec p\,^2)
  \sum_{m=0}^\infty\frac{iB_m}{E^2-m_m^2+i\epsilon}+\strut\nonumber\\&&\strut
  +\sum_{n=0}^\infty\frac{iB_n}{E^2-m_n^2+i\epsilon}(-i\vec p\,^2)
  \sum_{m=0}^\infty\frac{iB_m}{E^2-m_m^2+i\epsilon}(-i\vec p\,^2)
  \sum_{l=0}^\infty\frac{iB_l}{E^2-m_l^2+i\epsilon}+\ldots
  \ =\kern-2pt\nonumber\\
  &=&\sum_{n=0}^\infty\frac{iB_n}{E^2-m_n^2+i\epsilon}
  \left(1-\sum_{m=0}^\infty\frac{\vec p\,^2B_m}{E^2-m_m^2+i\epsilon}
  \right)^{-1}\ =\nonumber\\
  &=&\sum_{n=0}^\infty iB_n\left(E^2-m_n^2+i\epsilon-\vec p\,^2(E^2-m_n^2
  +i\epsilon)\sum_{m=0}^\infty\frac{B_m}{E^2-m_m^2+i\epsilon}\right)^{-1}.
\end{eqnarray}
Because of the sum rule, this restores the Lorentz invariance of the Green
function for high energies $E\gg m_0=\pi\omega/(2K(\kappa))$,
\begin{equation}
\tilde G_2(E;\vec p\,)=\sum_{n=0}^\infty
  \frac{iB_n}{E^2-\vec p\,^2-m_n^2+i\epsilon}=\sum_{n=0}^\infty
  \frac{iB_n}{p^2-m_n^2+i\epsilon}=\tilde G_2(p)
\end{equation}
in momentum space, and, therefore, the translational invariance in
configuration space. Finally, by considering the correlation function for the
scalar glueballs that is given by~\cite{Narison:2002woh,Narison:2021xhc}
\begin{equation}\label{eq:4corr}
{\cal O}(x)=\langle F^{a\mu\nu}(x)F^a_{\mu\nu}(x)
  F^{b\rho\eta}(0)F^b_{\rho\eta}(0)\rangle
\end{equation}
and whose poles are the physical glueballs. Using the technique explained
above and in Ref.~\cite{Frasca:2015yva}, one can see that according to
Ref.~\cite{Windisch:2012sz} the four-point correlator~(\ref{eq:4corr})
defining the correlation function of the glueball can be reduced to
convolutions over one- and two-point functions. As the one-point function has
no poles but zeros, the poles of the glueball four-point correlator are given
by the poles of the two-point correlator. Therefore, these poles represent
true colorless glueball states. Identifying the lowest glueball mass state
$m_0$ with the $\sigma$ resonance $f_0(500)$, one can fix the scale $k^2$ to
be $\sqrt{k^2}=\omega=2m_0K(-1)/\pi\approx 417\MeV$.

\end{appendix}


\begin{thebibliography}{99}

\bibitem{Witten:1985cc}
  E.~Witten,
  ``Noncommutative Geometry and String Field Theory,''
  Nucl.\ Phys.\ B \textbf{268}, 253--294 (1986)

\bibitem{Freund:1987kt}
  P.~G.~O.~Freund and M.~Olson,
  ``Nonarchimedean Strings,''
  Phys.\ Lett.\ B \textbf{199}, 186--190 (1987)

\bibitem{Freund:1987ck}
  P.~G.~O.~Freund and E.~Witten,
  ``Adelic String Amplitudes,''
  Phys.\ Lett.\ B \textbf{199}, 191 (1987)

\bibitem{Brekke:1987ptq}
  L.~Brekke, P.~G.~O.~Freund, M.~Olson and E.~Witten,
  ``Nonarchimedean String Dynamics,''
  Nucl.\ Phys.\ B \textbf{302}, 365--402 (1988)

\bibitem{Frampton:1988kr}
  P.~H.~Frampton and Y.~Okada,
  ``Effective Scalar Field Theory of $P^-$adic String,''
  Phys.\ Rev.\ D \textbf{37}, 3077--3079 (1988)

\bibitem{Kostelecky:1988ta}
  V.~A.~Kostelecky and S.~Samuel,
  ``The Static Tachyon Potential in the Open Bosonic String Theory,''
  Phys.\ Lett.\ B \textbf{207}, 169--173 (1988)

\bibitem{Kostelecky:1989nt}
  V.~A.~Kostelecky and S.~Samuel,
  ``On a Nonperturbative Vacuum for the Open Bosonic String,''
  Nucl.\ Phys.\ B \textbf{336}, 263--296 (1990)

\bibitem{Tseytlin:1995uq}
  A.~A.~Tseytlin,
  ``On singularities of spherically symmetric backgrounds in string theory,''
  Phys.\ Lett.\ B \textbf{363}, 223--229 (1995)

\bibitem{Siegel:2003vt}
  W.~Siegel,
  ``Stringy gravity at short distances,''
  [arXiv:hep-th/0309093 [hep-th]]

\bibitem{Biswas:2004qu}
  T.~Biswas, M.~Grisaru and W.~Siegel,
  ``Linear Regge trajectories from worldsheet lattice parton field theory,''
  Nucl.\ Phys.\ B \textbf{708}, 317--344 (2005)

\bibitem{Calcagni:2013eua}
  G.~Calcagni and L.~Modesto,
  ``Nonlocality in string theory,''
  J.\ Phys.\ A \textbf{47}, no.~35, 355402 (2014)

\bibitem{Calcagni:2014vxa}
  G.~Calcagni and L.~Modesto,
  ``Nonlocal quantum gravity and M-theory,''
  Phys.\ Rev.\ D \textbf{91}, no.~12, 124059 (2015)

\bibitem{Taylor:2003gn}
  W.~Taylor and B.~Zwiebach,
  ``D-branes, tachyons, and string field theory,''
  in ``Strings, Branes and Extra Dimensions,'' World Scientific,
  pp.~641--760 (2004)
  [arXiv:hep-th/0311017 [hep-th]].

\bibitem{Moffat:1990jj}
  J.~W.~Moffat,
  ``Finite nonlocal gauge field theory,''
  Phys.\ Rev.\ D \textbf{41}, 1177--1184 (1990)

\bibitem{Evens:1990wf}
  D.~Evens, J.~W.~Moffat, G.~Kleppe and R.~P.~Woodard,
  ``Nonlocal regularizations of gauge theories,''
  Phys.\ Rev.\ D \textbf{43}, no.~2, 499--519 (1991)

\bibitem{Kleppe:1991rv}
  G.~Kleppe and R.~P.~Woodard,
  ``Nonlocal Yang-Mills,''
  Nucl.\ Phys.\ B \textbf{388}, 81--112 (1992)

\bibitem{Tomboulis:1997gg}
  E.~T.~Tomboulis,
  ``Superrenormalizable gauge and gravitational theories,''
  [arXiv:hep-th/9702146 [hep-th]]

\bibitem{Clayton:2001xz}
  M.~A.~Clayton,
  ``Nonlocal regularization of Abelian models
  with spontaneous symmetry breaking,''
  Mod.\ Phys.\ Lett.\ A \textbf{16}, 1117--1127 (2001)

\bibitem{Biswas:2011ar}
  T.~Biswas, E.~Gerwick, T.~Koivisto and A.~Mazumdar,
  ``Towards singularity and ghost free theories of gravity,''
  Phys.\ Rev.\ Lett.\ \textbf{108}, 031101 (2012)

\bibitem{Modesto:2011kw}
  L.~Modesto,
  ``Super-renormalizable Quantum Gravity,''
  Phys.\ Rev.\ D \textbf{86}, 044005 (2012)

\bibitem{Modesto:2012ga}
  L.~Modesto,
  ``Super-renormalizable Higher-Derivative Quantum Gravity,''
  [arXiv:1202.0008 [hep-th]]

\bibitem{Modesto:2012ys}
  L.~Modesto,
  ``Super-renormalizable Multidimensional Quantum Gravity,''
  Astron.\ Rev.\ \textbf{8}, no.~2, 4--33 (2013)

\bibitem{Modesto:2013oma}
  L.~Modesto,
  ``Finite Quantum Gravity,''
  [arXiv:1305.6741 [hep-th]]

\bibitem{Biswas:2013cha}
  T.~Biswas, A.~Conroy, A.~S.~Koshelev and A.~Mazumdar,
  ``Generalized ghost-free quadratic curvature gravity,''
  Class.\ Quant.\ Grav.\ \textbf{31}, 015022 (2014)
  [erratum: Class.\ Quant.\ Grav.\ \textbf{31}, 159501 (2014)]

\bibitem{Modesto:2014xta}
  L.~Modesto,
  ``Multidimensional finite quantum gravity,''
  [arXiv:1402.6795 [hep-th]]

\bibitem{Modesto:2014lga}
  L.~Modesto and L.~Rachwa\l{},
  ``Super-renormalizable and finite gravitational theories,''
  Nucl.\ Phys.\ B \textbf{889}, 228--248 (2014)

\bibitem{Talaganis:2014ida}
  S.~Talaganis, T.~Biswas and A.~Mazumdar,
  ``Towards understanding the ultraviolet behavior of quantum loops
  in infinite-derivative theories of gravity,''
  Class.\ Quant.\ Grav.\ \textbf{32}, no.21, 215017 (2015)

\bibitem{Modesto:2015lna}
  L.~Modesto and L.~Rachwa\l{},
  ``Universally finite gravitational and gauge theories,''
  Nucl.\ Phys.\ B \textbf{900}, 147--169 (2015)

\bibitem{Modesto:2015foa}
  L.~Modesto, M.~Piva and L.~Rachwa\l{},
  ``Finite quantum gauge theories,''
  Phys.\ Rev.\ D \textbf{94}, no.~2, 025021 (2016)

\bibitem{Tomboulis:2015gfa}
  E.~T.~Tomboulis,
  ``Nonlocal and quasilocal field theories,''
  Phys.\ Rev.\ D \textbf{92}, no.12, 125037 (2015)

\bibitem{Modesto:2017hzl}
  L.~Modesto, L.~Rachwa\l{} and I.~L.~Shapiro,
  ``Renormalization group in super-renormalizable quantum gravity,''
  Eur.\ Phys.\ J.\ C \textbf{78}, no.7, 555 (2018)

\bibitem{Hashi:2018kag}
  M.~N.~Hashi, H.~Isono, T.~Noumi, G.~Shiu and P.~Soler,
  ``Higgs Mechanism in Nonlocal Field Theories,''
  JHEP \textbf{08}, 064 (2018)

\bibitem{Buoninfante:2018mre}
  L.~Buoninfante, G.~Lambiase and A.~Mazumdar,
  ``Ghost-free infinite derivative quantum field theory,''
  Nucl.\ Phys.\ B \textbf{944}, 114646 (2019)

\bibitem{Modesto:2021ief}
  L.~Modesto,
  ``Nonlocal Spacetime-Matter,''
  [arXiv:2103.04936 [gr-qc]]

\bibitem{Modesto:2021okr}
  L.~Modesto,
  ``The Higgs mechanism in nonlocal field theory,''
  JHEP \textbf{06}, 049 (2021)

\bibitem{Boos:2021chb}
  J.~Boos and C.~D.~Carone,
  ``Asymptotic nonlocality,''
  Phys.\ Rev.\ D \textbf{104}, no.~1, 015028 (2021)

\bibitem{Boos:2021jih}
  J.~Boos and C.~D.~Carone,
  ``Asymptotic nonlocality in gauge theories,''
  Phys.\ Rev.\ D \textbf{104}, no.~9, 095020 (2021)

\bibitem{Boos:2021lsj}
  J.~Boos and C.~D.~Carone,
  ``Asymptotic nonlocality in non-Abelian gauge theories,''
  Phys.\ Rev.\ D \textbf{105}, no.~3, 035034 (2022)

\bibitem{Boos:2022biz}
  J.~Boos and C.~D.~Carone,
  ``Asymptotically nonlocal gravity,''
  [arXiv:2212.00861 [hep-th]]

\bibitem{Krasnikov:1987yj}
  N.~V.~Krasnikov,
  ``Nonlocal Gauge Theories,''
  Theor.\ Math.\ Phys.\ \textbf{73}, 1184--1190 (1987)

\bibitem{Biswas:2014yia}
  T.~Biswas and N.~Okada,
  ``Towards LHC physics with nonlocal Standard Model,''
  Nucl.\ Phys.\ B \textbf{898}, 113--131 (2015)

\bibitem{Pius:2016jsl}
  R.~Pius and A.~Sen,
  ``Cutkosky rules for superstring field theory,''
  JHEP \textbf{10}, 024 (2016)
  [erratum: JHEP \textbf{09}, 122 (2018)]

\bibitem{Briscese:2018oyx}
  F.~Briscese and L.~Modesto,
  ``Cutkosky rules and perturbative unitarity
  in Euclidean nonlocal quantum field theories,''
  Phys.\ Rev.\ D \textbf{99}, no.~10, 104043 (2019)

\bibitem{Chin:2018puw}
  P.~Chin and E.~T.~Tomboulis,
  ``Nonlocal vertices and analyticity:
  Landau equations and general Cutkosky rule,''
  JHEP \textbf{06}, 014 (2018)

\bibitem{Briscese:2021mob}
  F.~Briscese and L.~Modesto,
  ``Non-unitarity of Minkowskian non-local quantum field theories,''
  Eur.\ Phys.\ J.\ C \textbf{81}, no.~8, 730 (2021)

\bibitem{Koshelev:2021orf}
  A.~S.~Koshelev and A.~Tokareva,
  ``Unitarity of Minkowski nonlocal theories made explicit,''
  Phys.\ Rev.\ D \textbf{104}, no.~2, 025016 (2021)

\bibitem{Ghoshal:2017egr}
  A.~Ghoshal, A.~Mazumdar, N.~Okada and D.~Villalba,
  ``Stability of infinite derivative Abelian Higgs models,''
  Phys.\ Rev.\ D \textbf{97}, no.~7, 076011 (2018)

\bibitem{Ghoshal:2020lfd}
  A.~Ghoshal, A.~Mazumdar, N.~Okada and D.~Villalba,
  ``Nonlocal non-Abelian gauge theory:
  Conformal invariance and \ensuremath{\beta}-function,''
  Phys.\ Rev.\ D \textbf{104}, no.~1, 015003 (2021)

\bibitem{Buoninfante:2018gce}
  L.~Buoninfante, A.~Ghoshal, G.~Lambiase and A.~Mazumdar,
  ``Transmutation of nonlocal scale in infinite derivative field theories,''
  Phys.\ Rev.\ D \textbf{99}, no.~4, 044032 (2019)

\bibitem{Ghoshal:2018gpq}
  A.~Ghoshal,
  ``Scalar dark matter probes the scale of nonlocality,''
  Int.\ J.\ Mod.\ Phys.\ A \textbf{34}, no.~24, 1950130 (2019)

\bibitem{Frasca:2022vvp}
  M.~Frasca, A.~Ghoshal and A.~S.~Koshelev,
  ``Quintessence dark energy from strongly-coupled higgs mass gap:
  local and non-local higher-derivative non-perturbative scenarios,''
  Eur.\ Phys.\ J.\ C \textbf{82}, no.~12, 1108 (2022)

\bibitem{Krasnikov:2020kgh}
  N.~V.~Krasnikov,
  ``Nonlocal SU(5) GUT,''
  Mod.\ Phys.\ Lett.\ A \textbf{36}, no.~15, 2150104 (2021)

\bibitem{Nortier:2021six}
  F.~Nortier,
  ``Shadow Extra Dimensions \& Fuzzy Branes with an Exorcized Spectrum,''
  [arXiv:2112.15592 [hep-th]]

\bibitem{Frasca:2020ojd}
  M.~Frasca and A.~Ghoshal,
  ``Diluted mass gap in strongly coupled non-local Yang-Mills,''
  JHEP \textbf{21}, 226 (2020)

\bibitem{Frasca:2021iip}
  M.~Frasca, A.~Ghoshal and N.~Okada,
  ``Confinement and renormalization group equations
  in string-inspired nonlocal gauge theories,''
  Phys.\ Rev.\ D \textbf{104}, no.~9, 096010 (2021)

\bibitem{Frasca:2020jbe}
  M.~Frasca and A.~Ghoshal,
  ``Mass gap in strongly coupled infinite derivative non-local Higgs:
  Dyson--Schwinger approach,''
  Class.\ Quant. Grav.\ \textbf{38}, no.~17, 17 (2021)

\bibitem{Frasca:2022duz}
  M.~Frasca, A.~Ghoshal and A.~S.~Koshelev,
  ``Non-perturbative Lee-Wick gauge theory:
  BRST, confinement \& RGE with strong couplings,''
  [arXiv:2202.09578 [hep-ph]]

\bibitem{Frasca:2022lwp}
  M.~Frasca, A.~Ghoshal and S.~Groote,
  ``Confinement in QCD and generic Yang-Mills theories
  with matter representations,''
  [arXiv:2202.14023 [hep-ph]]

\bibitem{Frasca:2022pjf}
  M.~Frasca, A.~Ghoshal and S.~Groote,
  ``Quark confinement in QCD in the 't Hooft limit,''
  Nucl.\ Part.\ Phys.\ Proc.\ \textbf{324--329}, 85--89 (2023)

\bibitem{Frolov:2015bia}
  V.~P.~Frolov, A.~Zelnikov and T.~de Paula Netto,
  ``Spherical collapse of small masses in the ghost-free gravity,''
  JHEP \textbf{06}, 107 (2015)

\bibitem{Frolov:2015usa}
  V.~P.~Frolov and A.~Zelnikov,
  ``Head-on collision of ultrarelativistic particles
  in ghost-free theories of gravity,''
  Phys.\ Rev.\ D \textbf{93}, no.~6, 064048 (2016)

\bibitem{Koshelev:2018hpt}
  A.~S.~Koshelev, J.~Marto and A.~Mazumdar,
  ``Schwarzschild $1/r$-singularity is not permissible
  in ghost free quadratic curvature infinite derivative gravity,''
  Phys.\ Rev.\ D \textbf{98}, no.~6, 064023 (2018)

\bibitem{Koshelev:2017bxd}
  A.~S.~Koshelev and A.~Mazumdar,
  ``Do massive compact objects without event horizon exist
  in infinite derivative gravity?,''
  Phys.\ Rev.\ D \textbf{96}, no.~8, 084069 (2017)

\bibitem{Buoninfante:2018xiw}
  L.~Buoninfante, A.~S.~Koshelev, G.~Lambiase and A.~Mazumdar,
  ``Classical properties of non-local, ghost- and singularity-free gravity,''
  JCAP \textbf{09}, 034 (2018)

\bibitem{Cornell:2017irh}
  A.~S.~Cornell, G.~Harmsen, G.~Lambiase and A.~Mazumdar,
  ``Rotating metric in nonsingular infinite derivative theories of gravity,''
  Phys.\ Rev.\ D \textbf{97}, no.~10, 104006 (2018)

\bibitem{Buoninfante:2018rlq}
  L.~Buoninfante, A.~S.~Koshelev, G.~Lambiase, J.~Marto and A.~Mazumdar,
  ``Conformally-flat, non-singular static metric
  in infinite derivative gravity,''
  JCAP \textbf{06}, 014 (2018)

\bibitem{Buoninfante:2018stt}
  L.~Buoninfante, G.~Harmsen, S.~Maheshwari and A.~Mazumdar,
  ``Nonsingular metric for an electrically charged point-source
  in ghost-free infinite derivative gravity,''
  Phys.\ Rev.\ D \textbf{98}, no.~8, 084009 (2018)

\bibitem{Abel:2019zou}
  S.~Abel, L.~Buoninfante and A.~Mazumdar,
  ``Nonlocal gravity with worldline inversion symmetry,''
  JHEP \textbf{01}, 003 (2020)

\bibitem{Buoninfante:2020ctr}
  L.~Buoninfante, G.~Lambiase, Y.~Miyashita, W.~Takebe and M.~Yamaguchi,
  ``Generalized ghost-free propagators in nonlocal field theories,''
  Phys.\ Rev.\ D \textbf{101}, no.~8, 084019 (2020)

\bibitem{Biswas:2005qr}
  T.~Biswas, A.~Mazumdar and W.~Siegel,
  ``Bouncing universes in string-inspired gravity,''
  JCAP \textbf{03}, 009 (2006)

\bibitem{Biswas:2006bs}
  T.~Biswas, R.~Brandenberger, A.~Mazumdar and W.~Siegel,
  ``Non-perturbative Gravity, Hagedorn Bounce \&\ CMB,''
  JCAP \textbf{12}, 011 (2007)

\bibitem{Biswas:2010zk}
  T.~Biswas, T.~Koivisto and A.~Mazumdar,
  ``Towards a resolution of the cosmological singularity
  in non-local higher derivative theories of gravity,''
  JCAP \textbf{11}, 008 (2010)

\bibitem{Biswas:2012bp}
  T.~Biswas, A.~S.~Koshelev, A.~Mazumdar and S.~Y.~Vernov,
  ``Stable bounce and inflation in non-local higher derivative cosmology,''
  JCAP \textbf{08}, 024 (2012)

\bibitem{Koshelev:2012qn}
  A.~S.~Koshelev and S.~Y.~Vernov,
  ``On bouncing solutions in non-local gravity,''
  Phys.\ Part.\ Nucl.\ \textbf{43}, 666--668 (2012)

\bibitem{Koshelev:2018rau}
  A.~S.~Koshelev, J.~Marto and A.~Mazumdar,
  ``Towards resolution of anisotropic cosmological singularity
  in infinite derivative gravity,''
  JCAP \textbf{02}, 020 (2019)

\bibitem{Kolar:2021qox}
  I.~Kol\'a\v{r}, F.~J.~M.~Torralba and A.~Mazumdar,
  ``New nonsingular cosmological solution of nonlocal gravity,''
  Phys.\ Rev.\ D \textbf{105}, no.~4, 044045 (2022)

\bibitem{Ghoshal:2022mnj}
  A.~Ghoshal and F.~Nortier,
  ``Fate of the false vacuum in string-inspired nonlocal field theory,''
  JCAP \textbf{08}, no.~08, 047 (2022)

\bibitem{Boos:2018bxf}
  J.~Boos, V.~P.~Frolov and A.~Zelnikov,
  ``Gravitational field of static p-branes in linearized ghost-free gravity,''
  Phys.\ Rev.\ D \textbf{97}, no.~8, 084021 (2018)

\bibitem{Boos:2018bhd}
  J.~Boos,
  ``Gravitational Friedel oscillations in higher-derivative and
  infinite-derivative gravity?,''
  Int.\ J.\ Mod.\ Phys.\ D \textbf{27}, no.~14, 1847022 (2018)

\bibitem{Boos:2020kgj}
  J.~Boos,
  ``Angle deficit and nonlocal gravitoelectromagnetism around a slowly
  spinning cosmic string,''
  Int.\ J.\ Mod.\ Phys.\ D \textbf{29}, no.~14, 2043027 (2020)

\bibitem{Boos:2020qgg}
  J.~Boos,
  ``Effects of Non-locality in Gravity and Quantum Theory,''
  [arXiv:2009.10856 [gr-qc]]

\bibitem{Lee:1969fy}
  T.~D.~Lee and G.~C.~Wick,
  ``Negative Metric and the Unitarity of the S Matrix,''
  Nucl.\ Phys.\ B \textbf{9}, 209-243 (1969)

\bibitem{Lee:1970iw}
  T.~D.~Lee and G.~C.~Wick,
  ``Finite Theory of Quantum Electrodynamics,''
  Phys.\ Rev.\ D \textbf{2}, 1033-1048 (1970)

\bibitem{Gabrielli:2007cp}
  E.~Gabrielli,
  ``On the dynamical breaking of chiral symmetry: A New mechanism,''
  Phys.\ Rev.\ D \textbf{77}, 055020 (2008)

\bibitem{Bender:1999ek} 
  C.~M.~Bender, K.~A.~Milton and V.~M.~Savage,
  ``Solution of Schwinger-Dyson equations for PT
  symmetric quantum field theory,''
  Phys.\ Rev.\ D {\bf 62}, 085001 (2000)

\bibitem{Frasca:2007uz}
  M.~Frasca,
  ``Infrared Gluon and Ghost Propagators,''
  Phys.\ Lett.\ B {\bf 670}, 73--77 (2008)

\bibitem{Frasca:2009yp}
  M.~Frasca,
  ``Mapping a Massless Scalar Field Theory on a Yang-Mills Theory:
  Classical Case,''
  Mod.\ Phys.\ Lett.\ A {\bf 24}, 2425--2432 (2009)

\bibitem{Smilga:2001ck}
  A.~V.~Smilga,
  ``Lectures on quantum chromodynamics,''
  World Scientific, Singapore, 2001

\bibitem{Matinyan:1981ys}
  S.~G.~Matinyan, G.~K.~Savvidy and N.~G.~Ter-Arutunian Savvidy,
  ``Stochasticity of Classical {Yang-Mills} Mechanics and Its Elimination
  by Higgs Mechanism. (In Russian),''
  JETP Lett.\ {\bf 34}, 590--593 (1981)

\bibitem{Savvidy:1982wx}
  G.~K.~Savvidy,
  ``Yang--Mills Classical Mechanics as a Kolmogorov K System,''
  Phys.\ Lett.\ B {\bf 130}, 303--307 (1983)

\bibitem{Savvidy:1982jk}
  G.~K.~Savvidy,
  ``Classical and Quantum Mechanics of Nonabelian Gauge Fields,''
  Nucl.\ Phys.\ B {\bf 246}, 302--334 (1984)

\bibitem{Athenodorou:2021qvs}
  A.~Athenodorou and M.~Teper,
  ``SU(N) gauge theories in 3+1 dimensions:
  glueball spectrum, string tensions and topology,''
  JHEP \textbf{12}, 082 (2021)

\bibitem{Lucini:2004my}
  B.~Lucini, M.~Teper and U.~Wenger,
  ``Glueballs and k-strings in $SU(N)$ gauge theories: Calculations with
  improved operators,''
  JHEP \textbf{06}, 012 (2004)

\bibitem{Frasca:2017mrh}
  M.~Frasca,
  ``Spectrum of a four-dimensional Yang-Mills theory,''
  [arXiv:1705.07647 [hep-ph]]

\bibitem{Frasca:2017slg} 
  M.~Frasca,
  ``Spectrum of Yang-Mills theory in 3 and 4 dimensions,''
  Nucl.\ Part.\ Phys.\ Proc.\ \textbf{294--296}, 124 (2018)

\bibitem{Chernodub:2023dok}
  M.~N.~Chernodub, V.~A.~Goy, A.~V.~Molochkov and A.~S.~Tanashkin,
  ``Boundary states and Non-Abelian Casimir effect
  in lattice Yang-Mills theory,''
  [arXiv:2302.00376 [hep-lat]]

\bibitem{Frasca:2011fj}
  M.~Frasca,
  ``Casimir effect in Yang-Mills theories,''
  [arXiv:1108.6299 [hep-ph]]

\bibitem{Hell:2008cc}
  T.~Hell, S.~R\"o\ss ner, M.~Cristoforetti and W.~Weise,
  ``Dynamics and thermodynamics of a non-local Polyakov--Nambu--Jona-Lasinio
  model with running coupling,''
  Phys.\ Rev.\ D {\bf 79}, 014022 (2009)

\bibitem{Nambu:1961tp}
  Y.~Nambu and G.~Jona-Lasinio,
  ``Dynamical Model of Elementary Particles Based on an
  Analogy with Superconductivity. 1.,''
  Phys.\ Rev.\  {\bf 122}, 345 (1961)

\bibitem{Frasca:2013tma}
  M.~Frasca,
  ``Scalar field theory in the strong self-interaction limit,''
  Eur.\ Phys.\ J.\ C {\bf 74}, 2929 (2014)

\bibitem{Frasca:2015yva}
  M.~Frasca,
  ``Quantum Yang-Mills field theory,''
  Eur.\ Phys.\ J.\ Plus {\bf 132}, 38 (2017)
  [erratum: Eur.\ Phys.\ J.\ Plus {\bf 132}E, 242 (2017)]

\bibitem{Abbott:1981ke}
  L.~F.~Abbott,
  ``Introduction to the Background Field Method,''
  Acta Phys.\ Polon.\ B {\bf 13}, 33 (1982)

\bibitem{Narison:2002woh}
  S.~Narison,
  ``QCD as a Theory of Hadrons: From Partons to Confinement,''
  Camb.\ Monogr.\ Part.\ Phys.\ Nucl.\ Phys.\ Cosmol.\ \textbf{17},
  1-812 (2007).

\bibitem{Narison:2021xhc}
  S.~Narison,
  ``Di-gluonium sum rules, $I=0$ scalar mesons and conformal anomaly,''
  Nucl.\ Phys.\ A \textbf{1017}, 122337 (2022).

\bibitem{Windisch:2012sz}
  A.~Windisch, M.~Q.~Huber and R.~Alkofer,
  ``On the analytic structure of scalar glueball operators at the Born level,''
  Phys.\ Rev.\ D \textbf{87}, no.6, 065005 (2013).

\end{thebibliography}
\end{document}